%% file: paper_arxiv.tex
\documentclass[aps, prd, reprint,10pt, notitlepage, a4paper,floats, floaqutfix,amsmath, amssymb, amsfonts,superscriptaddress,showpacs, showkeys,nofootinbib,longbibliography]{revtex4-1}

\pdfoutput=1

\usepackage{epsfig}
\usepackage{graphics}
\usepackage{graphicx}
\usepackage{amsmath,amssymb,mathrsfs}
\usepackage{amsfonts}
\usepackage[usenames,dvipsnames]{xcolor}
\usepackage{wasysym}
\usepackage{times}
\usepackage{mathptmx}
\usepackage{gensymb}
\usepackage{appendix}
\usepackage{listings}
\usepackage{url}
\usepackage[normalem]{ulem}
\usepackage{alltt}
\usepackage[colorlinks]{hyperref}
\usepackage{cleveref}
\usepackage{longtable}
\usepackage{enumitem}
\setlist{nosep}
\usepackage{color}
\usepackage{calc}
\usepackage{tensor}
\usepackage{bm}
\usepackage{times}
\usepackage{multirow}
\usepackage[varg]{txfonts}
\usepackage{float}
\usepackage{dcolumn}
\usepackage[nolist,nohyperlinks]{acronym}
\usepackage{xspace}
\usepackage[english]{babel}
\usepackage[abs]{overpic}
\usepackage{pict2e}
\usepackage[caption=false]{subfig} 
\allowdisplaybreaks[1]
\usepackage[utf8]{inputenc}
\usepackage{gensymb}
\usepackage{bm}
\usepackage{stackengine}
\usepackage{boldline,multirow}
\usepackage{braket}
\usepackage{longtable}

\usepackage{tabularx}
\usepackage{rotating}

\DeclareMathAlphabet{\mathcalstd}{OMS}{cmsy}{m}{n}
\DeclareMathAlphabet{\mathpzc}{OT1}{pzc}{m}{it}

\newcommand{\AEI}{Max Planck Institute for Gravitational Physics (Albert Einstein Institute), Am M{\"u}hlenberg 1, Potsdam, 14476, Germany}
\newcommand{\Maryland}{Department of Physics, University of Maryland, College Park, MD 20742, USA}

\definecolor{dodgerblue}{HTML}{1E90FF}
\definecolor{viennared}{HTML}{DA0A14}

\hypersetup{citecolor=dodgerblue}

\acrodef{PN}{post-Newtonian}
\acrodef{EOB}{effective-one-body}
\acrodef{NR}{numerical relativity}
\acrodef{GW}{gravitational wave}
\acrodef{BBH}{binary black hole}
\acrodef{BH}{black hole}
\acrodef{BNS}{binary neutron star}
\acrodef{NSBH}{neutron star-black hole}
\acrodef{SNR}{signal-to-noise ratio}
\acrodef{aLIGO}{Advanced LIGO}
\acrodef{AdV}{Advanced Virgo}

\begin{document}



\title{Effective-one-body multipolar waveforms for eccentric binary black holes with non-precessing spins}

\author{Antoni Ramos-Buades}
\affiliation{\AEI}
\author{Alessandra Buonanno}
\affiliation{\AEI}
\affiliation{\Maryland}
\author{Mohammed Khalil}
\affiliation{\AEI}
\affiliation{\Maryland}
\author{Serguei Ossokine}
\affiliation{\AEI}

\date{\today}

\begin{abstract}
We construct an inspiral-merger-ringdown eccentric gravitational-wave (GW) 
model for binary black holes with non-precessing spins within the
effective-one-body formalism. This waveform model, \texttt{SEOBNRv4EHM},  
extends the accurate quasi-circular \texttt{SEOBNRv4HM} model 
to eccentric binaries by including recently computed eccentric corrections up 
to 2PN order in the gravitational waveform modes, 
notably the $(l,|m|)=(2,2),(2,1),(3,3),(4,4),(5,5)$ multipoles. The waveform model
reproduces the zero eccentricity limit with an accuracy comparable to the
underlying quasi-circular model, with the unfaithfulness of $\lesssim 1 \%$ against
quasi-circular numerical-relativity (NR) simulations. When compared against 28 public
eccentric NR simulations  from the Simulating eXtreme Spacetimes catalog  with initial orbital eccentricities
up to $e\simeq 0.3$ and dimensionless spin magnitudes up to $+0.7$, the model
provides unfaithfulness $< 1\%$,  showing that both the $(2,|2|)$-modes
and the higher-order modes are reliably described without calibration 
to NR datasets in the eccentric sector. The waveform model \texttt{SEOBNRv4EHM} is able 
to qualitatively reproduce the phenomenology of dynamical captures, 
and can be extended to include spin-precession effects. It can be employed for 
upcoming observing runs with the LIGO-Virgo-KAGRA detectors and used 
to re-analyze existing GW catalogs to infer the eccentricity parameters for binaries with 
$ e \lesssim 0.3$ (at 20 Hz or lower) and spins up to $\lesssim 0.9-0.95$. The latter is 
a promising region of the parameter space where some astrophysical formation 
scenarios of binaries predict mild eccentricity in the ground-based detectors' bandwidth. 
Assessing the accuracy and robustness of the eccentric waveform model \texttt{SEOBNRv4EHM} for 
larger eccentricities and spins will require comparisons with, and, likely, calibration to eccentric NR waveforms in 
a larger region of the parameter space.
\end{abstract}

\maketitle


\section{Introduction}\label{sec:introduction}
\label{sec:intro}

Most inspiraling binaries observed by ground-based gravitational-wave (GW)
detectors are likely to form via isolated binary evolution \cite{Bethe:1998bn,Belczynski:2001uc, Dominik:2013tma, Belczynski:2014iua, mennekens2014massive,spera2015mass, Belczynski:2016obo, Eldridge:2016ymr, Marchant:2016wow,Mapelli:2017hqk,Mapelli:2018wys,Stevenson:2017tfq,Giacobbo:2018etu,Kruckow:2018slo,Kruckow:2018slo}
 and are expected to circularize~\cite{Peters:1964zz} 
by the time they enter the detector frequency band. 
However, a small fraction of binaries
may have non-negligible orbital eccentricity in the LIGO, 
Virgo  or KAGRA \cite{TheLIGOScientific:2014jea,TheVirgo:2014hva,KAGRA:2018plz} 
frequency band if they form through dynamical captures and interactions in dense stellar environments, such as globular clusters~\cite{PortegiesZwart:1999nm, Miller:2001ez,Miller:2002pg,Gultekin:2004pm,Gultekin:2005fd,OLeary:2005vqo, Sadowski:2007dz, downing2010compact,downing2011compact, Samsing:2013kua,Rodriguez:2015oxa,Askar:2016jwt,Rodriguez:2016kxx, Rodriguez:2016avt, Samsing:2017rat,Samsing:2017xmd,Rodriguez:2017pec,Rodriguez:2018rmd, Fragione:2018vty, Zevin:2018kzq,Gondan:2020svr} or galactic nuclei~\cite{OLeary:2008myb,Antonini:2012ad,Tsang:2013mca,Antonini:2016gqe,Petrovich:2017otm,Stone:2016wzz,Stone:2016ryd,Rasskazov:2019gjw}, and through the Kozai-Lidov mechanism \cite{Kozai:1962zz,Lidov:1962zz} in triple systems~\cite{Wen:2002km,VanLandingham:2016ccd,Rodriguez:2016vmx,Antonini:2017ash,Fragione:2019dtr,Fragione:2018yrb,Fragione:2019hqt}. Thus, measuring eccentricity in the GW signal from merging binaries provides key information about the origin and the properties of the population of such binaries \cite{Mandel:2009nx,LIGOScientific:2016vpg, Farr:2017uvj, LIGOScientific:2018jsj, Zevin:2021rtf,LIGOScientific:2021psn}.

So far, the observed GW events detected by LIGO and Virgo~\cite{LIGOScientific:2018mvr, LIGOScientific:2020ibl,LIGOScientific:2021djp} are consistent with quasi-circular binary coalescences. Nevertheless, there are increasing efforts to search for eccentricity signatures in the current GW events~\cite{LIGOScientific:2019dag,Romero-Shaw:2019itr,Nitz:2019spj,
Romero-Shaw:2020thy,Gayathri:2020coq,Favata:2021vhw,
OShea:2021ugg, Romero-Shaw:2021ual}. With upcoming upgrades of ground-based
detectors and third-generation detectors like the Einstein Telescope or the Cosmic Explorer \cite{Punturo:2010zz,LIGOScientific:2016wof,Reitze:2019dyk,
Reitze:2019iox}, as well as future space-borne detectors like LISA and TianQin
\cite{amaroseoane2017laser,TianQin:2015yph}, the fraction of GW events with
non-negligible orbital eccentricity is expected to significantly increase
\cite{Sesana:2010qb,Breivik:2016ddj,Samsing:2018isx,Cardoso:2020iji}. Therefore,
developing accurate  waveform models that include the effects of eccentricity is
essential to detect eccentric binaries, infer their properties, and provide
information on their astrophysical origin.

Gravitational waveforms from inspiraling eccentric binaries have been
developed within the post-Newtonian (PN)
formalism~\cite{Gopakumar:1997bs, Gopakumar:2001dy, Damour:2004bz,
  Konigsdorffer:2006zt, Arun:2007rg,Arun:2007sg, Arun:2009mc,
  Memmesheimer:2004cv,Yunes:2009yz,Huerta:2014eca,Mishra:2015bqa,
  Loutrel:2017fgu,Klein:2018ybm,Moore:2018kvz,Moore:2019xkm,Tanay:2019knc,Tiwari:2020hsu}. 
Numerical-relativity (NR) simulations for eccentric binary black holes
(BBHs) were produced in
Refs.~\cite{Hinder:2008kv,Huerta:2019oxn,Lewis:2016lgx,Habib:2019cui,Ramos-Buades:2019uvh,Gayathri:2020coq,Islam:2021mha}, but they are still limited to a small region of the binary's parameter
space and do not cover the entire bandwidth of ground-based detectors
(unless the binary total mass is larger than $\sim 70 M_\odot$
\cite{Islam:2021mha}). Under the
assumption that the binary circularizes before merger,
inspiral-merger-ringdown (IMR) (hybrid) waveforms, in time or
frequency-domain, have been developed in
Refs.~\cite{Hinder:2017sxy,Huerta:2017kez,Ramos-Buades:2019uvh} by
combining the inspiral phase from PN with the merger and ringdown
signal from either NR or the effective-one-body (EOB) formalism.
Recently, NR surrogate models for equal-mass non-spinning eccentric
binaries were built in Refs.~\cite{Islam:2021mha} by directly
interpolating NR simulations. Guided by comparisons with NR
simulations, Ref.~\cite{Setyawati:2021gom} has proposed a method to
include eccentricity effects in existing quasi-circular IMR waveform
models for low eccentricity. Regarding systems with matter content,
like binary neutron stars or neutron-star--black-hole binaries, there
have been also efforts to produce eccentric NR
simulations~\cite{Gold:2011df,East:2011xa,Yang:2018bzx}, as well as
analytical work studying the coupling between eccentricity and tidal
effects~\cite{Chirenti:2016xys,Samsing:2016bqm,Yang:2019kmf}.
However, complete eccentric IMR waveform models including matter effects have not been developed, yet.

Within the efforts to model IMR waveforms for eccentric BBHs, the EOB formalism \cite{Buonanno:1998gg,Buonanno:2000ef} has recently
seen a lot of progress \cite{Hinderer:2017jcs,Cao:2017ndf, Liu:2019jpg,Chiaramello:2020ehz,Liu:2021pkr,Nagar:2021gss,Nagar:2021xnh, Albanesi:2021rby,
Khalil:2021txt,Placidi:2021rkh}. The EOB formalism is a framework that combines information from PN theory, NR and BH perturbation theory 
to accurately describe the inspiral, merger and  ringdown of a binary coalescence (see e.g., Refs.~\cite{Damour:2008gu,Pan:2010hz,Pan:2013rra,
Taracchini:2013rva,Bohe:2016gbl,Nagar:2018zoe,Cotesta:2018fcv,Babak:2016tgq,Nagar:2018gnk,Ossokine:2020kjp,Riemenschneider:2021ppj,Mihaylov:2021bpf,Nagar:2021xnh,Gamba:2021ydi}). The current 
eccentric EOB waveform models are constructed by improving the EOB description of the eccentric inspiral and plunge, but they still employ 
a quasi-circular merger-ringdown model~\cite{Liu:2021pkr, Chiaramello:2020ehz,Nagar:2021gss,Nagar:2021xnh,Placidi:2021rkh}. Nevertheless, this approach 
has been able to construct EOB waveforms that are faithful to existing, although limited, (public) NR waveforms  from the Simulating eXtreme Spacetimes
  (\texttt{SXS}) catalog with eccentricity smaller than 
$~0.3$ and mild spins.

In this paper, we develop a multipolar eccentric EOB waveform model
that builds on the quasi-circular \texttt{SEOBNRv4HM}
model~\cite{Cotesta:2018fcv} for BBHs with aligned spins\footnote{To ease the notation we use the term aligned spins when referring to aligned/anti-aligned spins.}
and includes recently derived eccentric corrections up to 2PN
  order~\cite{Khalil:2021txt}, including spin-orbit and spin-spin
  interactions, in the $(l,|m|)=(2,2),(2,1),(3,3),(4,4),(5,5)$
  multipoles. This eccentric waveform model, henceforth
  \texttt{SEOBNRv4EHM}, has comparable accuracy to the quasi-circular
  \texttt{SEOBNRv4HM} model in the zero eccentricity limit when
  compared to quasi-circular NR waveforms, and produces
  unfaithfulness $< 1\%$ against eccentric NR
  simulations~\cite{Hinder:2017sxy}  from the \texttt{SXS} collaboration.  When restricting to
  the $(2,|2|)$-modes we refer to the model as \texttt{SEOBNRv4E}, in
  analogy to the quasi-circular case, which corresponds to the
  \texttt{SEOBNRv4} model in Ref.~\cite{Bohe:2016gbl}. Furthermore, we
  develop generic initial conditions for elliptical orbits including
  two eccentric parameters. We also implement hyperbolic-orbit initial
  conditions, and we briefly show the ability of the model to
  reproduce the phenomenology of hyperbolic encounters, thus paving
  the path to the description of generic BBH coalescences.

  This paper is structured as follows. In
  Sec.~\ref{sec:EccEOBexpressions}, we outline our multipolar
  eccentric EOB waveforms, and describe how the eccentricity
  effects are introduced in each building block of the model,  notably the 
conservative and dissipative dynamics and the gravitational waveform modes. 
We also develop initial conditions for elliptic orbits including two eccentric
  parameters. In Sec.~\ref{sec:WaveformValidation}, we assess the accuracy of the
  multipolar eccentric waveform model by comparing it against 141 NR
  waveforms in the quasi-circular limit, and to 28 public eccentric NR
  waveforms from the \texttt{SXS} waveform catalog
  \cite{Boyle:2019kee,SXS:catalog}. We develop an algorithm to
  estimate the best matching parameters between the eccentric EOB and
  NR waveforms, analyze the robustness of the model across
  parameter space and start to estimate for which source's parameters and 
eccentricity, we could anticipate biases in inference studies 
if quasi-circular--orbit waveforms were used. In Sec.~\ref{sec:conclusions}, we summarize our
  main conclusions and discuss future work. Finally, in Appendix \ref{app:EccModes} we list the eccentric corrections to the waveform modes obtained in Ref.~\cite{Khalil:2021txt}, in Appendix \ref{sec:AppendixA} we describe details of the implementation of the eccentric waveforms modes, and in Appendix \ref{app:eccICs} we provide the expressions of the dynamical quantities  needed for calculating the initial conditions for eccentric orbits.
  
In this paper, we use geometric units, setting $G=c=1$ unless otherwise specified.

\section{Eccentric effective-one-body  waveform model}\label{sec:EccEOBexpressions}

Here, we develop the multipolar eccentric aligned-spin \texttt{SEOBNRv4EHM} waveform model 
building on the quasi-circular aligned-spin \texttt{SEOBNRv4HM} model~\cite{Cotesta:2018fcv}, 
which has been used by LIGO and Virgo to detect GW signals and infer binary properties~\cite{LIGOScientific:2018mvr,
LIGOScientific:2020ibl,LIGOScientific:2021djp}. More specifically, we provide a brief description of the (conservative) dynamics in Sec.~\ref{sec:RRforces}, 
waveform modes in Sec.~\ref{sec:WaveformModes}, and initial conditions in Sec.~\ref{sec:InitialConditions}.

The EOB formalism maps the two-body dynamics of objects with masses $m_i$
and spins $\bm{S}_i$, with $i=1,2$, into an effective dynamics of a test-spin with mass $\mu=m_1m_2/(m_1+m_2)$ and spin $\bm{S}_*$ moving 
in a deformed Kerr metric with mass $M=m_1+m_2$ and spin $\bm{S}_{\text{Kerr}}$. The deformation parameter is the (dimensionless)  
symmetric mass ratio $\nu = \mu/M$. As we are limiting to spins aligned to the orbital angular momentum, the only (dimensionless) 
spin component on which the dynamics and the waveform depend is $\chi_i= \bm{S}_i \cdot \hat{\bm{L}}/m_i^2$, where $\hat{\bm{L}}$ 
is the unit vector in the direction perpendicular to the orbital plane.

\subsection{Effective-one-body dynamics}  \label{sec:RRforces} 

The EOB conservative dynamics is governed by the
EOB Hamiltonian, calculated from the effective Hamiltonian through the energy map~\cite{Buonanno:1998gg}
\begin{equation}
H_{\text{EOB}}=M \sqrt{1+ 2 \nu \left(\frac{H_{\text{eff}}}{\mu}-1 \right) }\,. 
\label{eq:eq0}
\end{equation}
When both spins are aligned with the orbital angular momentum, the motion is
restricted to a plane. This implies that the dynamical variables entering the
Hamiltonian are the (dimensionless) radial separation $r \equiv R/M$, the orbital phase $\phi$, and their
(dimensionless) conjugate momenta $p_r \equiv P_r/\mu$ and $p_\phi \equiv P_\phi / \mu$. We use the same effective Hamiltonian,
$H_{\text{eff}}$, as described in Refs.~\cite{Barausse:2011ys,Taracchini:2013rva}, augmented with the 
parameters $(K,d_{\text{SO}}, d_{\text{SS}}, \Delta t^{22}_{\text{peak}})$ calibrated to NR waveforms from Ref.~\cite{Bohe:2016gbl}.

The dissipative dynamics within the EOB formalism is described by a radiation-reaction (RR) force $\mathcal{F}$, which 
enters the Hamilton equations of motion, as~\cite{Pan:2011gk,Pan:2013rra}
\begin{align}
\dot{r} &= \xi(r) \frac{\partial \hat{H}_\text{EOB}}{\partial p_{r_*}}(r,p_{r_*},p_\phi), \nonumber\\
\dot{\phi} &= \frac{\partial \hat{H}_\text{EOB}}{\partial p_\phi}(r,p_{r_*},p_\phi), \nonumber\\
\dot{p}_{r_*} &= - \xi(r) \frac{\partial \hat{H}_\text{EOB}}{\partial r}(r,p_{r_*},p_\phi) + \hat{\mathcal{F}}_r, \nonumber\\
\dot{p}_\phi &= \hat{\mathcal{F}}_\phi,
\label{eq:eq1}
\end{align}
where the dot represents the time derivative $d/d\hat{t}$, with respect to the dimensionless time $\hat{t} \equiv T/M$, $\hat{H}_\text{EOB}\equiv H_\text{EOB}/\mu$, and $\hat{\mathcal{F}}_\phi \equiv \mathcal{F}_\phi / M$.
The equations are expressed in terms of $p_{r_*} \equiv p_r\,\xi(r)$, which is the conjugate momentum to the tortoise-coordinate $r_*$, and $\xi(r)\equiv dr/dr_*$ can be expressed in terms of the potentials of the effective Hamiltonian~\cite{Pan:2011gk}.

In the case of the \texttt{SEOBNRv4HM} waveform model, the components of the RR force are computed using the following relations~\cite{Buonanno:2000ef,Buonanno:2005xu}
\begin{equation}
\hat{\mathcal{F}}_\phi = - \frac{\Phi_E}{\omega}, \quad  \hat{\mathcal{F}}_r=\hat{\mathcal{F}}_\phi \frac{p_r}{p_\phi}, \\
\label{eq:eq2}
\end{equation}
where $\omega = \dot{\phi}$ is the (dimensionless) orbital frequency, and $\Phi_E$ is the energy flux for quasi-circular orbits written as a sum over waveform modes using~\cite{Damour:2008gu,Pan:2010hz}
\begin{equation}
\Phi_E = \frac{\omega^2}{16\pi} \sum_{l=2}^{8} \sum_{m=-l}^{l} m^2 \left| \frac{D_L}{M} h_{lm}\right|^2,
\label{eq:eq2.1}
\end{equation}
where $D_L$ is the luminosity distance between the binary system and the observer. The above relation is only valid for quasi-circular orbits as it assumes 
the relation between energy and angular-momentum fluxes $\Phi_E = \omega \Phi_J$, which is only valid for quasi-circular orbits.

We note that in the \texttt{SEOBNRv4HM} model, eccentric effects are already partially included in the radial component of the RR force $\hat{\mathcal{F}}_r$ 
since it is proportional to $p_r$, whereas the tangential component of the RR force $\hat{\mathcal{F}}_\phi$ does not contain eccentric corrections. 
Recently, Ref.~\cite{Khalil:2021txt} derived the eccentric corrections of the RR force up to 2PN order, 
including spin-orbit and spin-spin interactions, in a factorized form~\cite{Damour:2008gu,Pan:2010hz}. We have explored adding those corrections to both components 
of the RR force in the \texttt{SEOBNRv4HM} model. However, we find, when doing it, that the late-inspiral dynamics can lead to differences with respect 
to the one of the \texttt{SEOBNRv4HM} model, affecting the inclusion of the merger-ringdown signal that is inherited from the \texttt{SEOBNRv4HM} model. 
This in turn, can lead to differences between our new model and \texttt{SEOBNRv4HM} in the quasi-circular orbit limit, and, for some binary configurations, 
to the degradation of the model performance when compared to quasi-circular NR simulations. Since the goal of this paper is to develop an eccentric 
waveform model that reduces to the \texttt{SEOBNRv4HM} model in the quasi-circular limit and  is faithful to the current (public) SXS 
NR eccentric waveforms (which have eccentricity smaller than $0.3$), we choose  to retain the conservative and dissipative dynamics
of the \texttt{SEOBNRv4HM} model and introduce the eccentric corrections of Ref.~\cite{Khalil:2021txt} only in the gravitational modes.
The latter are not used to compute the fluxes employed to construct the RR force. We leave the inclusion of the eccentric corrections 
to the RR force for the next generation of EOBNR models~\cite{seobnrv5}, which will be recalibrated to quasi-circular NR simulations\footnote{The quasi-circular \texttt{TEOBResumS} model~\cite{Nagar:2018zoe} does not include the radial component of the RR force $\hat{\mathcal{F}}_r$, 
but its extension to eccentric orbits~\cite{Nagar:2021gss} includes a non-zero $\hat{\mathcal{F}}_r$ that is linear in $p_r$, adds eccentric 
corrections at Newtonian order in $\hat{\mathcal{F}}_\phi$, and is recalibrated to NR waveforms in the quasi-circular orbit limit.}.

\subsection{Effective-one-body gravitational waveforms}
\label{sec:WaveformModes}

As in previous EOBNR models, we represent the inspiral-plunge signal of the \texttt{SEOBNRv4EHM} waveforms as:
\begin{equation}
h_{lm}^\text{insp-plunge} = h^{\text{ecc}}_{lm}\, N_{lm},
\end{equation}
where the $h^{\text{ecc}}_{lm}$'s are the factorized EOB gravitational modes~\cite{Damour:2008gu,Pan:2010hz}, including the 2PN eccentric corrections derived in Ref.~\cite{Khalil:2021txt}, while the $N_{lm}$'s are the so-called nonquasi-circular (NQC) terms. More specifically, the $h^{\text{ecc}}_{lm}$ terms are written as:
\begin{equation}
h^{\text{ecc}}_{lm} = h^{\text{N}}_{lm}\, S_{\text{eff}}\, (T^{\text{qc}}_{lm}+T^{\text{ecc}}_{lm})\,(f^{\text{qc}}_{lm}+f^{\text{ecc}}_{lm})\, e^{i \delta_{lm}},
\label{eq:eq5}
\end{equation}
where $h^{\text{N}}_{lm}$ is the Newtonian (leading-order) quasi-circular (qc) term, $S_{\text{eff}}$ is an effective source term, $T^{\text{qc}}_{lm}$ resums the leading logarithms in the tail effects, while $\delta_{lm}$ contains phase corrections, and $f^{\text{qc}}_{lm}$ ensures that the PN expansion of $h^{\text{qc}}_{lm}$  agrees with the PN expressions for the modes in the quasi-circular orbit limit. The explicit expressions for the above terms can be found in Refs.~\cite{Pan:2010hz,Taracchini:2012ig,Bohe:2016gbl,Cotesta:2018fcv}. Furthermore, the term $T^{\text{ecc}}_{lm}$ includes eccentric corrections to the leading-order hereditary part, while  $f^{\text{ecc}}_{lm}$ contains the eccentric corrections to the 2PN instantaneous part, including the Newtonian (leading-order) term. 
Note that the eccentric corrections are not introduced in $\delta_{\ell m}$. We note that the eccentric-orbit terms are provided in the Supplemental Material of Ref.~\cite{Khalil:2021txt}, 
and for completeness, we write them in Appendix~\ref{app:EccModes} for the $(l,|m|)=\{(2,2),(2,1),(3,3),(4,4),(5,5)\}$ modes.

We use the same expression of the NQC correction as in the \texttt{SEOBNRv4HM} model, that is~\cite{Bohe:2016gbl,Cotesta:2018fcv}
\begin{equation}
\begin{split}
N_{lm} & = \left[ 1+ \frac{p^2_{r_*}}{(r \omega)^2}\left( a_1^{lm} + \frac{a_2^{lm}}{r}  + \frac{a_3^{lm}}{r^{3/2}} \right) \right] \\
&  \times \exp \left[i \left( b_1^{lm} \frac{p_{r_*}}{r \omega} + b_2^{lm} \frac{p^3_{r_*}}{r \omega}  \right)  \right] ,
\end{split}
\label{eq:eq6}
\end{equation}
where the coefficients $(a^{lm}_1,a^{lm}_2,a^{lm}_3,b_1^{lm}, b_2^{lm})$ are
fixed by requiring that the amplitude, its first and second
derivatives, as well as the GW frequency and its first derivative  agree for every
$(l,m)$-mode with values extracted from NR waveforms~\cite{Cotesta:2018fcv}  (i.e., the NR input values). However, as we shall discuss below, we 
orbit-average the NQC corrections (\ref{eq:eq6}) in the eccentric \texttt{SEOBNRv4HM} waveform model.


Both the eccentric corrections to the waveform and the NQC terms are
designed to improve the accuracy of the EOB waveforms. However, we find that modifications to
both have to be introduced in order to improve the faithfulness of the EOB
model to NR simulations. The eccentric corrections to the modes are derived
from PN and EOB theory, thus they increase the accuracy of the inspiral part of the
eccentric model. Nonetheless, in the strong-field regime, very close to the
merger-ringdown attachment time \cite{Bohe:2016gbl,Cotesta:2018fcv}, we found
that they can lead to high unfaithfulness with respect to NR waveforms. This is
due to the fact that they can modify by orders of magnitude the NR input values (in
particular the amplitude, frequency and their derivatives) used to compute the
coefficients in the NQC terms. To mitigate this effect, we introduce a
sigmoid function that makes the eccentric corrections, $f^{\text{ecc}}_{lm}$ and $T^{\text{ecc}}_{lm}$, vanish at merger,
\begin{equation}
w(\beta,t_\beta;t) = \frac{1}{1+e^{-\beta (t-t_\beta)}},
\label{eq:eq7}
\end{equation}
where we choose $\beta=0.09$ and $t_\beta \equiv t^\omega_{\text{peak}}-300$, being $t^\omega_{\text{peak}}$ the time when the peak of $\omega\equiv \dot{\phi}$ occurs. 

Moreover, the NQC corrections to the waveform defined in Eq.
\eqref{eq:eq6} become highly oscillatory during an eccentric
inspiral, as all the dynamical quantities composing its ansatz have increasing
oscillations with increasing eccentricity. 
One approach to circumvent the oscillatory behaviour of the NQC function is the
application of a window function \cite{Liu:2021pkr,
Nagar:2021gss,Nagar:2021xnh,Placidi:2021rkh},  like the sigmoid in Eq. \eqref{eq:eq7}, close to
merger. This window function forces the NQC function to approach unity during
the inspiral, and have significant effects only near merger.

Here we develop an alternative approach, which consists in orbit averaging the
dynamical quantities entering the ansatz of the NQC corrections, so that it is a monotonic
function during the whole evolution. 
The rationale is that the genuine oscillations due to the orbital eccentricity are already included 
in Eq. (\ref{eq:eq5}), thus the role of the NQC corrections is merely to improve the 
GW amplitude and frequency of the EOB model during plunge and merger using inputs from NR.

Any dynamical quantity $X(t)$ can be orbit-averaged at the $i^\text{th}$-orbit passage as
follows,
\begin{equation}
\overline{X}_i =  \frac{1}{t_{i+1}-t_{i}} \int^{t_{i+1}}_{t_{i}} X_i(t) dt,
\label{eq:eq9}
\end{equation}
where $(t_i,t_{i+1})$ correspond to times defining the complete orbits. 
We note that in order to perform the orbit average calculation, we
need to identify the times, $t_i$, which define successive orbits. In
practice, we can choose either the maxima or the minima of any orbital
quantity to identify the orbits. In the case of the orbital
separation, these times correspond to the turning points, either apastron or
periastron passages. We decide to use the time of the
maxima to perform the orbit average in Eq. \eqref{eq:eq9}, and we
associate each orbit average value to an intermediate time defined as
$\overline{t}_i=(t_{i+1}+t_i)/2$ \cite{Lewis:2016lgx}. From 
Eq. \eqref{eq:eq6} it can be seen that the dynamical quantities
entering the NQC corrections are $r(t)$, ${p}_{r_*}(t)$ and $\omega(t)$. Thus, the NQC
corrections implemented in the model can be expressed in terms of the
orbit-average quantities as,
\begin{equation}
\begin{split}
\overline{N}_{lm} & = \left[ 1+ \frac{\overline{p}_{r_*}^2}{\overline{r}^2 \overline{\omega}^2}\left( a_1^{lm} + \frac{a_2^{lm}}{\overline{r} }  + \frac{a_3^{lm}}{\overline{r}^{3/2}} \right) \right] \\
&  \times \exp \left[i \left( b_1^{lm} \frac{\overline{p}_{r_*}}{\overline{r}\,\overline{\omega} } + b_2^{lm} \frac{\overline{p}_{r_*}^3}{\overline{r}\, \overline{\omega}}  \right)  \right] ,
\end{split}
\label{eq:eq10}
\end{equation}
where the coefficients $(a^{lm}_1,a^{lm}_2,a^{lm}_3,b_1^{lm}, b_2^{lm})$ are computed as in Eq. \eqref{eq:eq6}. Hence, the modes in \texttt{SEOBNRv4EHM} can be expressed as follows,
\begin{align}
\overline{h}_{lm}^\text{insp-plunge} & = h^{\text{N}}_{lm}\,S_{\text{eff}}\, T_{lm}\, f_{lm}\, e^{i \delta_{lm}}\, \overline{N}_{lm}, \\
T_{lm} &=   T^{\text{qc}}_{lm}+\left[1-w(\beta,t_\beta;t)\right]T^{\text{ecc}}_{lm}, \\
f_{lm} &=  f^{\text{qc}}_{lm}+\left[1-w(\beta,t_\beta;t)\right] f^{\text{ecc}}_{lm}.
\label{eq:eq11}
\end{align}
where the NQC terms $\overline{N}_{lm}$ are given by Eq. \eqref{eq:eq10}. The full details
of the  orbit averaging procedure  can be found in the Appendix~\ref{sec:AppendixA}.

The procedure described above ensures that during an eccentric inspiral there
are no unphysical oscillations coming from the oscillatory nature of the
dynamical variables in the NQC correction, while the windowing applied to the eccentric terms close 
to merger ensures that the circularization
hypothesis at merger is fulfilled, making the input values 
of the eccentric model closer to the ones of the underlying quasi-circular model. In
Sec.~\ref{sec:Robustness}, we quantify the validity in parameter space of the
approximations used to treat the NQC corrections. 
Furthermore, in Secs.~\ref{sec:NRquasicircular} and \ref{sec:NReccentric}, we show 
that this procedure provides a quasi-circular limit with an accuracy comparable 
to the underlying quasi-circular \texttt{SEOBNRv4HM} model, and a high faithfulness 
when compared to eccentric NR simulations.

\subsection{Eccentric initial conditions}  \label{sec:InitialConditions}

We now complete the eccentric waveform model with the specification
of the initial conditions for elliptical orbits and hyperbolic orbits.

The gravitational signal emitted by an aligned-spin eccentric BBH
system is described by 6 intrinsic 
parameters: the component masses $m_1$ and $m_2$ (or equivalently mass
ratio $q=m_1/m_2$ and total mass $M=m_1+m_2$), the dimensionless spin
components $\chi_1$ and $\chi_2$ introduced at
  the beginning of Sec.~\ref{sec:EccEOBexpressions}, the orbital
eccentricity $e$, and a radial phase parameter $\zeta$. For the 
  parameter describing the position of a point on an ellipse, several
  options with different physical meaning are possible: mean anomaly,
  relativistic anomaly, true anomaly, etc.. Here, we adopt the
  relativistic anomaly. In General Relativity, for BBH systems, the total mass
is just a scale parameter that can be set to 1. Thus, the initial conditions for the EOB evolution of the
\texttt{SEOBNRv4EHM} model depend only on 5 parameters, which have to be specified at a certain
starting frequency $\omega_0$.

Since the eccentricity parameter $e$ is gauge dependent, we can choose a measure
of the eccentricity that is as convenient as possible for the numerical
implementation. The only requirement is that, in the zero eccentricity
limit, we recover the quasi-circular initial
conditions~\cite{Buonanno:2005xu} used in the \texttt{SEOBNRv4HM}
model. Reference~\cite{Khalil:2021txt} derived such initial conditions for
eccentric orbits assuming the periastron as the starting point. Here,
we generalize those initial conditions to start from an arbitrary
point on the orbit, thus making the eccentric initial conditions depend on both $e$ and $\zeta$.

We use the eccentricity $e$ defined in the Keplerian parametrization of the orbit
\begin{equation}
r = \frac{1}{u_p (1 + e \cos \zeta)} \,,
\end{equation}
where $u_p$ is the inverse semilatus rectum, and $\zeta$ the relativistic anomaly, which equals $0$ at periastron and $\pi$ at apastron.
Given the initial orbital frequency $\omega_0$, eccentricity $e_0$, relativistic anomaly $\zeta_0$, masses, and spins, we obtain the initial conditions for $r_0$ and ${p_\phi}_0$ in absence of radiation reaction by solving the following equations:
\begin{equation}
\left[\frac{\partial \hat{H}_\text{EOB}}{\partial r}\right]_0 = - \left[\dot p_r(p_\phi,e,\zeta)\right]_0, \qquad \left[\frac{\partial \hat{H}_\text{EOB}}{\partial p_\phi}\right]_0 = \omega_0,
\end{equation}
with $p_r(p_\phi,e,\zeta)$ and $\dot p_r(p_\phi,e,\zeta)$ given by the 2PN-order expressions given
in Eqs.~\eqref{pr2PN} and~\eqref{prdot2PN} of the Appendix~\ref{app:eccICs}.

Using the solution for $r_0$ and ${p_\phi}_0$, we obtain the initial condition for $p_{r_0}$ by numerically solving
\begin{equation}
\left[\frac{\partial \hat{H}_\text{EOB}}{\partial p_r}\right]_0 = \left[\dot{r}^{(0)} + \dot{r}^{(1)}\right]_0,
\end{equation}
where $\dot{r}^{(0)}$ is the 2PN-order expression for $\dot{r}$ at zeroth order in the RR effects (see Eq.~\eqref{rdot2PN}), while $\dot{r}^{(1)}$ is the first-order term in the RR part of $\dot{r}$, for which we use the 
quasi-circular expression derived in Ref.~\cite{Buonanno:2005xu}
\begin{equation}
\dot{r}^{(1)} = - \frac{\Phi_E^\text{qc}}{\omega} \frac{\partial^2 \hat{H}_\text{EOB} / \partial r\partial p_\phi}{\partial^2 \hat{H}_\text{EOB} / \partial r^2},
\end{equation}
being $\Phi_E^\text{qc}$ the quasi-circular energy flux given in Eq.~\eqref{eq:eq2.1}. Finally, the initial value  $p_{r_0}$ is converted into the tortoise-coordinate conjugate momentum ${p_{r_*}}_0$, using the relations 
in Sec. \ref{sec:RRforces}, so that together with $r_0$ and ${p_\phi}_0$, it can be introduced in Eqs. \eqref{eq:eq1} to  evolve the EOB equations of motion. 

It is worth to compare our initial orbital eccentricity 
$e^{\text{IC}}_0$, with the eccentricity measured directly from the orbital
frequency, $e_{\omega_{\text{orb}}}$, and the eccentricity computed from the frequency of
the $(2,2)$ mode, $e_{\omega_{22}}$, by using the following eccentricity
estimator~\cite{PhysRevD.66.101501,Ramos-Buades:2019uvh}
\begin{equation}
e_\omega = \frac{\omega^{1/2}_p-\omega^{1/2}_a}{\omega^{1/2}_p+\omega^{1/2}_a},
\label{eq:eq13}
\end{equation}
 where $\omega_a$ and $\omega_p$ correspond to the frequency, either the orbital or $(2,2)$-mode frequency, at apastron and periastron, respectively.

\begin{figure}[htbp!]
\includegraphics[width= \columnwidth]{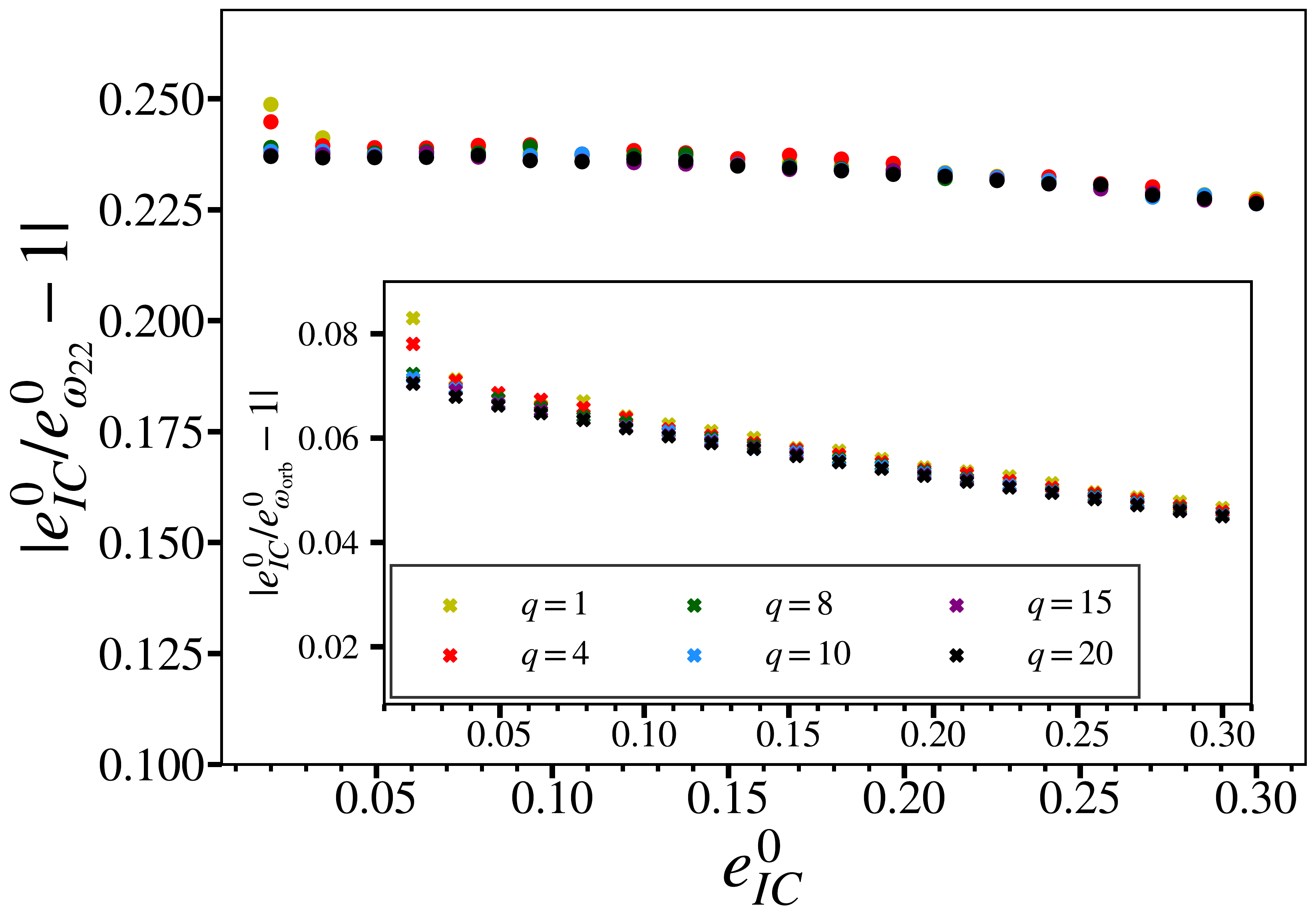}
\includegraphics[width= \columnwidth]{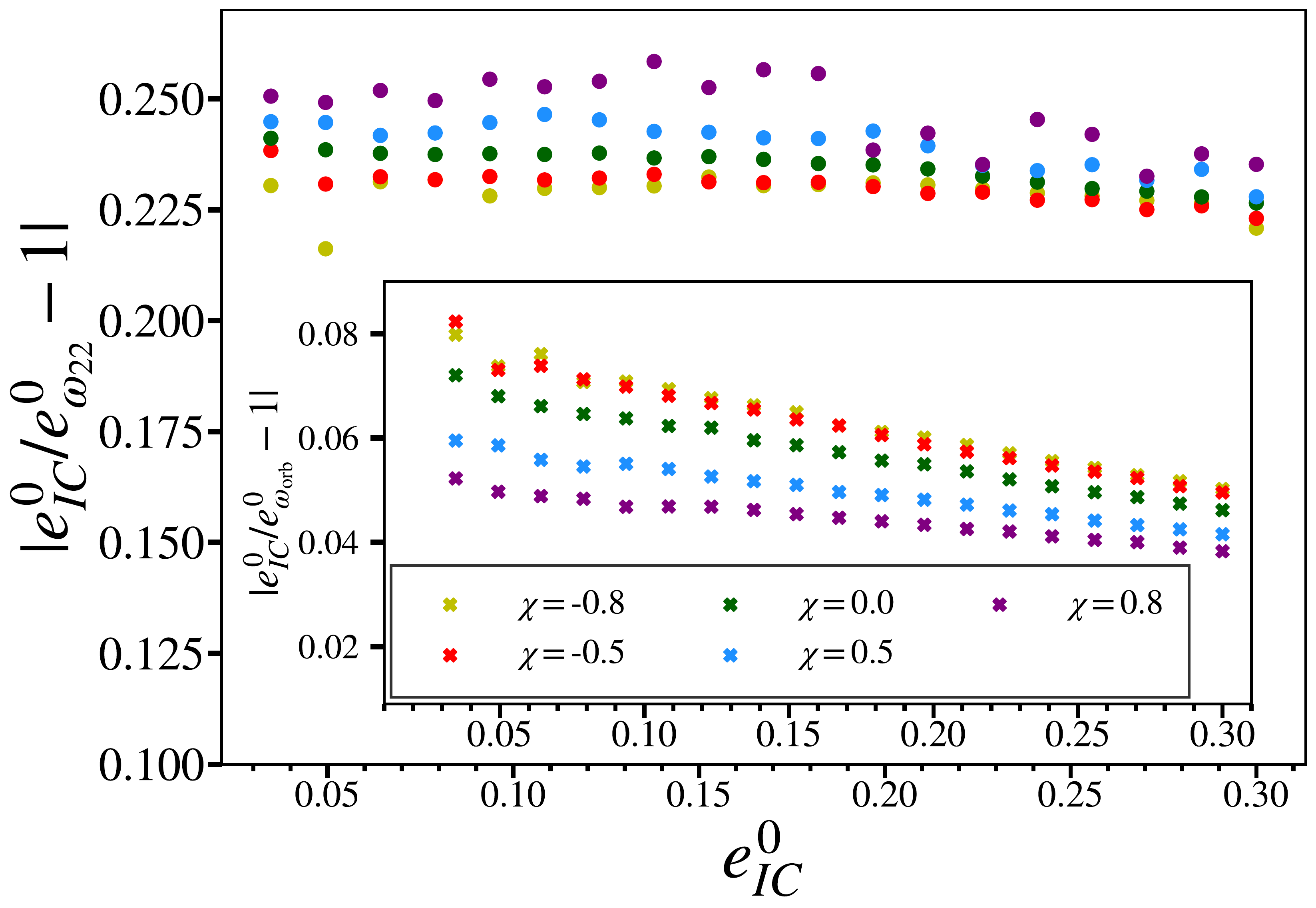}
\caption{Top panel: Relative difference between the initial eccentricity from the initial conditions of \texttt{SEOBNRv4EHM}, $e^0_{IC}$, and the initial eccentricity  computed from the  frequency of the $(2,2)$ mode,  $e^0_{\omega_{22}}$, of \texttt{SEOBNRv4EHM} as a function of $e^0_{IC}$ in the range $[0.02,0.3]$ for non-spinning configurations with mass ratios $q=\{1,4,8,10,15,20\}$. Lower panel: Same quantity as in the upper panel for a configuration with fixed mass ratio $q=2$, same range of $e^0_{IC}$, and distinct equal-spin values $\chi_1 =\chi_2 = \{-0.8,-0.5,0,0.5,0.8\}$. The insets of both panels show the relative difference between  $e^0_{IC}$ and the eccentricity measured from the orbital frequency of \texttt{SEOBNRv4EHM} $e^0_{\omega_{\text{orb}}}$, for the same configurations as in the larger panels.}
\label{fig:initialConditions}
\end{figure}

Starting at periastron ($\zeta=0$), we produce a sample of $5\times
10^5$ points randomly distributed in the parameter space $q\in
[1,20]$, $\chi_{1,2}\in[-0.9,0.9]$ and $e^{\text{IC}}_0 \in
[0.01,0.3]$, and compute the relative difference between $e^0_{IC}$ and
$e^0_{\omega_{22}}$ or $e^0_{\omega_{\text{orb}}}$.  In the upper
panel of Fig. \ref{fig:initialConditions}, we show the relative  difference 
between $e^{\text{IC}}_0$ and $e^0_{\omega_{22}}$ for some
non-spinning configurations with mass ratios $
q=\{1,4,8,10,15,20\}$. For the same cases we show in the inset of
Fig. \ref{fig:initialConditions} the relative difference  between $e^0_{IC}$
and $e^0_{\omega_{\text{orb}}}$. We observe that the relative  difference 
for $e^0_{\omega_{22}}$ is $\sim 24 \%$, while for
$e^0_{\omega_{\text{orb}}}$ is significantly lower $\sim
6\%$. Moreover, the dependence on mass ratio is smaller than $1 \%$, in
relative difference, with the exception of the lower eccentricity cases where
the measurement of the eccentricity has also a larger error, which we
estimate to be $\sim 3 \%$. We also note that with increasing values of 
$e^0_{IC}$ the relative differences decrease, especially for the
eccentricity measured from the orbital frequency. 

\begin{figure*}
\begin{center}
\includegraphics[scale=0.25]{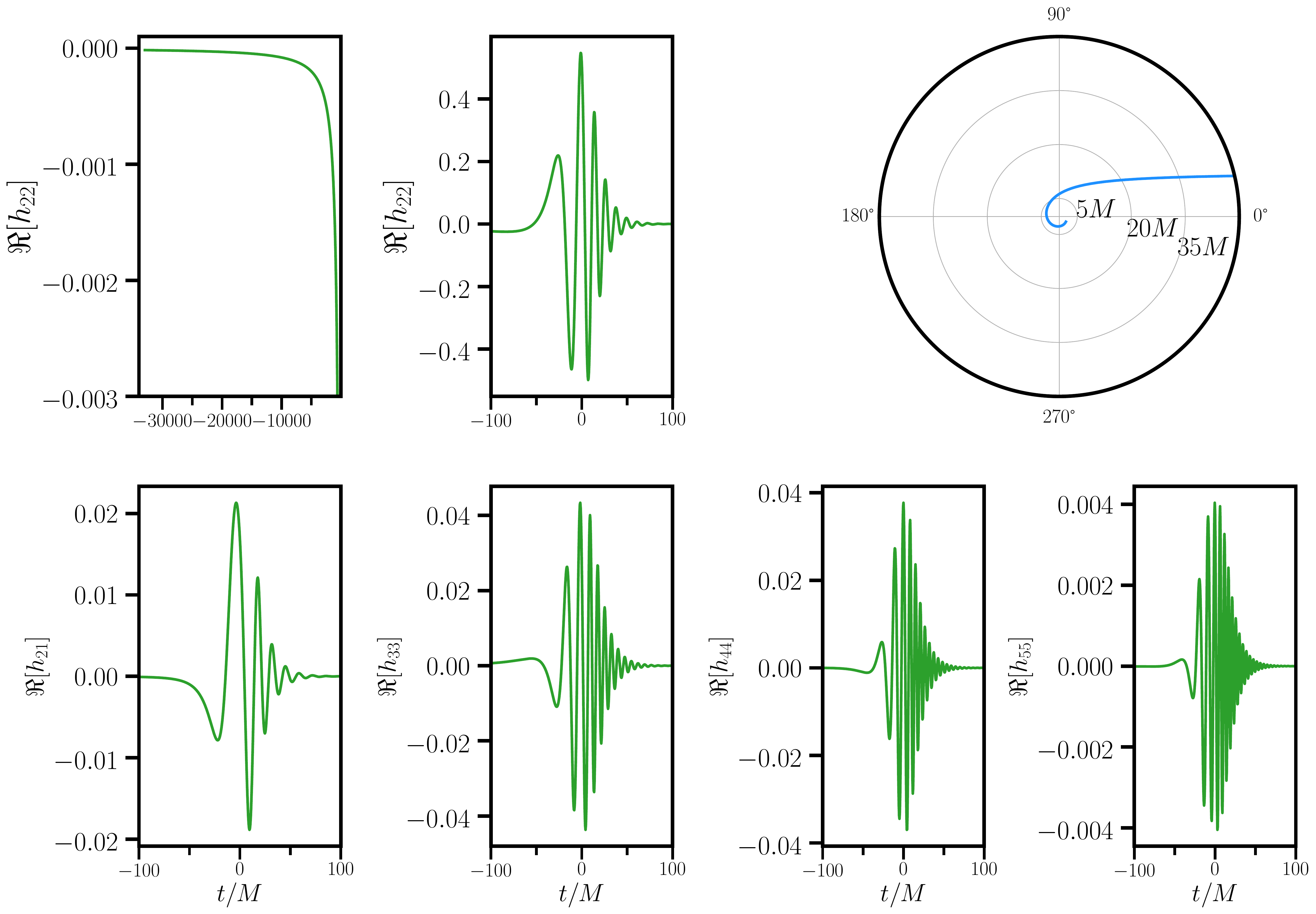}
\caption{Waveform characteristics for a mass ratio $q=1.5$ non-spinning configuration with initial parameters $r_0=10^4M$, ${p_{\phi}}_0=3.97$ and $E_0/M=1.012$. \textit{Top row:} From left to right, the two first panels show the real part of the 22-mode waveform in time domain, the first plot displays the full domain of the waveform while the second one zooms in into the merger part. The third panel displays the trajectory $r(\varphi)$ in polar coordinates. Each circle corresponds to a constant value of the orbital separation, which is marked on the figure. 
 \textit{Bottom row:} From left to right, the real part of the $(2,1),(3,3),(4,4),(5,5)$ modes in time domain zooming in close to the merger and ringdown regions.}
\label{fig:hypEncounter}
\end{center}

\end{figure*}

In the lower panel of Fig.~\ref{fig:initialConditions}, we fix the mass ratio $q=2$, and
vary the spin values $\chi_1 =\chi_2 = \{-0.8,-0.5,0,0.5,0.8\}$ for the same
range of initial eccentricities as in the upper panel. The results
show that the relative error between $e^0_{\omega_{22}}$ and
$e^0_{IC}$ is $\sim 22\mbox{--}25 \% $ when varying the spin values. These variations are
quite similar to the results obtained when varying the mass ratios. The
relative error between $e^0_{\omega_{\text{orb}}}$ and $e^0_{IC}$ is
$\sim 5\mbox{--}8\%$, which is also very close in magnitude to the
non-spinning case. However, in the lower panel of
Fig.~\ref{fig:initialConditions}, one can appreciate that the relative
errors vary more with spins than with mass ratio, specifically for
positive spins the relative errors with respect to $e^0_{\omega_{22}}$
are $1\mbox{--}2\%$ larger than for negative spins, while the relative errors
for $e^0_{\omega_{\text{orb}}}$ follow the inverse dependence with
spins.

When considering the larger dataset of $5\times 10^5$ configurations,
we find that the relative error between $e^{\text{IC}}_0$ and
$e^0_{\omega_{22}}$ has an average of $\sim 30 \%$ error, with the
largest difference of $40\%$ for low eccentricities, where the errors
in measuring the eccentricity are also larger due to the difficulties
in determining the maxima and minima. For the relative error between
$e^{\text{IC}}_0$ and $e^0_{\omega_{\text{orb}}}$, the average value
is $\sim 6\%$, reaching up to $\sim 14 \%$ for low values of the
eccentricities.  The results show a better agreement between
$e^0_{IC}$ and $e^0_{\omega_{\text{orb}}}$ than between
$e^0_{IC}$ and $e^0_{\omega_{22}}$. The relation between these different definitions
of eccentricity can be derived using PN theory, and it will be
presented in future work \cite{eccGSFNR}.

We note that the results reported in Fig.~\ref{fig:initialConditions} 
quantify differences between the eccentricity specified in the \texttt{SEOBNRv4EHM} model and two other possible definitions of the eccentricity. We remark that even though
there is no unique definition of the eccentricity, this kind of quantitative
analysis will be required in future parameter-estimation analysis of eccentric GW sources with the LIGO, Virgo 
and KAGRA detectors in order to reliably compare the results among different eccentric waveform models.

Finally, we have also implemented hyperbolic-orbit initial conditions for the
\texttt{SEOBNRv4EHM} model. The hyperbolic initial conditions are specified by 
the initial energy $E_0$ and angular momentum ${p_{\phi}}_0$ at infinity, which in practice
we take at an initial orbital separation $r_0 = 10^4M$. We fix a value of the angular momentum ${p_{\phi}}_0$, and choose a value of the initial energy  $E_0/M$. The choice of $E_0/M$ is typically 
done between the energy with zero radial momentum $E_{\text{min}}=H_\text{EOB}(r_0,{p_\phi}_0,{p_{r_*}}_0=0)$ and the energy at the last stable circular orbit (LSO)
$E_{\text{max}}=H_\text{EOB}(r_{\text{LSO}},{p_\phi}^\text{LSO},{p_{r^*}^\text{LSO}})$~\cite{Damour:2014afa,Nagar:2020xsk,Gamba:2021gap}. Then, we solve the following equation for $p_{{r_*}_0}$, 
\begin{equation}
E_0 \equiv H_\text{EOB}(r_0,{p_\phi}_0,{p_{r_*}}_0).
\end{equation}
This procedure to set the initial conditions for hyperbolic orbits 
is very similar to the one used in the literature~\cite{Damour:2014afa,Nagar:2020xsk}, although we note that others are possible,
for instance,  one could express the initial conditions in terms of the initial velocity and impact parameter
\cite{Bini:2017pee,Cho:2018upo,Mukherjee:2020hnm}. In Fig.~\ref{fig:hypEncounter}, we show the trajectory
of a dynamical capture, as well as, the real part of the different multipoles in the \texttt{SEOBNRv4EHM} model. Although the model is able to reproduce the behavior of dynamical
captures and hyperbolic encounters, we focus in this paper on the eccentric
bound orbit case, and leave a thorough and quantitative analysis of the
hyperbolic orbits, including comparison to NR results to the future.

\section{Performance of the multipolar eccentric effective-one-body waveform model}\label{sec:WaveformValidation}


In this section, we first assess the accuracy of the multipolar eccentric waveform model \texttt{SEOBNRv4EHM} against the quasi-circular 
and eccentric NR waveforms at our disposal, using the faithfulness function, which is 
a metric introduced to quantify the closeness of two waveforms. Then, we explore the robustness and validity of 
the \texttt{SEOBNRv4EHM} model in the region of parameter space where we do not yet have NR waveforms. Finally, we evaluate the unfaithfulness between IMR eccentric waveforms and quasi-circular ones to estimate in which part of the parameter space and for which values of the eccentricity we expect large biases in recovering the source's properties if quasi-circular orbit waveforms were used.

\subsection{Faithfulness function}
\label{sec:NRunfaithfulness}

The GW signal emitted by an eccentric aligned-spin BBH
system depends on 13 parameters. In Sec.~\ref{sec:InitialConditions}, we have introduced the
$6$ intrinsic parameters describing the source properties of such a
system. There are 7 additional parameters which
relate the source and detector frames, notably the angular position of the line of 
sight measured in the source frame ($\iota$, $\varphi_0$), the sky location of the source 
in the detector frame $(\theta,\phi)$, the polarization angle $\psi$, the luminosity distance 
of the source $D_L$ and the time of arrival $t_c$.

The signal measured by the detector takes the form:
\begin{equation}
\begin{split}
h(t) = & F_+(\theta,\phi,\psi) h_+(\iota, \varphi_0,D_L,\bm{\Theta},t_c;t)\\
&+ F_\times(\theta,\phi,\psi) h_\times(\iota, \varphi_0,D_L,\bm{\Theta},t_c;t),
\end{split}
\label{eq:eq14}
\end{equation}
where $\bm{\Theta}=\{m_{1,2},\chi_{1,2},e,\zeta\}$, and $F_+(\theta,\phi,\psi) $ and $F_\times(\theta,\phi,\psi)$ are the antenna-pattern functions \cite{PhysRevD.44.3819,Finn:1992xs}. Equation~\eqref{eq:eq14} can be written in terms of an effective polarization angle $\kappa (\theta,\phi,\psi)$ as
\begin{equation}
h(t) =  \mathcal{A}(\theta,\phi)\left[ h_+ \cos \kappa + h_+ \sin \kappa\right],
\label{eq:eq14.1}
\end{equation}
where the definition of $\mathcal{A}(\theta,\phi)$ can be found in Refs.~\cite{Cotesta:2018fcv,Ossokine:2020kjp}, and we have removed the dependences of $\kappa$, $h_+$ and $h_\times$ to ease the notation.
The GW polarizations can be decomposed as
\begin{equation}
h_+-ih_\times =  \sum^{\infty}_{l=2} \sum_{m=-l}^{m=+l}{}_{-2} Y_{lm}(\varphi,\iota)\, h_{lm}(\bm{\Theta};t),
\label{eq:eq14.2}
\end{equation}
where $h_{lm}(\bm{\Theta};t)$ represents the gravitational waveform modes, and $ Y^{-2}_{lm}(\varphi,\iota)$ are the -2 spin-weighted spherical harmonics.

We introduce the inner product between two waveforms $h_1$ and $h_2$ \cite{PhysRevD.44.3819,Finn:1992xs} as
\begin{equation}
\langle h_1 | h_2 \rangle  = 4 \Re \int^{f_{\text{max}}}_{f_{\text{min}}} \frac{\hat{h}_1(f)\hat{h}^*_2(f)}{S_n(f)}df,
\label{eq:eq15}
\end{equation}
where the star denotes complex conjugate, the hat the Fourier transform, and
$S_n(f)$ is the one-sided power-spectral density (PSD) of the detector noise. In
this work we use the Advanced LIGO's zero-detuned high-power design
sensitivity curve \cite{Barsotti:2018}. When both waveforms are in band, we use
$f_{\text{min}}=10$Hz and $f_{\text{max}}=2048$Hz, as the lower and upper bounds
of the integral. For NR waveforms where this is not the case, we set $f_{\text{min}} = 1.05 f_{\text{start}}$, where $f_{\text{start}}$ is the starting frequency of the NR waveform.

The agreement between two waveforms --- for example, the signal, $h_s$, and the template, $h_t$, observed by a detector, can be assessed by computing the faithfulness
function \cite{Cotesta:2018fcv,Ossokine:2020kjp},
\begin{equation}
\label{eq:eq16}
\mathcal{F}(M_{\textrm{s}},\iota_{\textrm{s}},{\varphi_0}_{\textrm{s}},\kappa_{\textrm{s}}) =  \max_{t_c, {\varphi_0}_{t}, \kappa_{t}} \left[\left . \frac{ \langle h_s|h_t \rangle}{\sqrt{  \langle h_s|h_s \rangle  \langle h_t|h_t \rangle}}\right \vert_{\substack{\iota_{\mathrm{s}} = \iota_{t} \\\boldsymbol{\Theta}_\mathrm{s}(t_{\mathrm{s}} = t_{0_\mathrm{s}}) = \boldsymbol{\Theta}_{t}(t_t = t_{0_\mathrm{t}})}} \right ].
\end{equation}
In Eq. \eqref{eq:eq16} the inclination angle of the signal and the template are
set to be the same, while the coalescence time, azimuthal angle and effective
polarization angle of the template $(t_{0_t},\varphi_{0_t}, \kappa_t)$, are
adjusted to maximize the faithfulness of the template. This is a typical choice
made when comparing waveforms with higher-order modes~\cite{Cotesta:2018fcv,Ossokine:2020kjp,Garcia-Quiros:2020qpx}. It is
convenient to introduce the \textit{sky-and-polarization averaged faithfulness} to reduce the dimensionality of the faithfulness function and
express it in a more compact form \cite{Cotesta:2018fcv,Ossokine:2020kjp},
\begin{equation}
\label{eq:eq17}
\overline{ \mathcal{F}}(M_{\textrm{s}}) =  \frac{1}{8 \pi^2}\int^{1}_{-1} d{(\cos \iota_{\textrm{s}})} \int^{2 \pi}_0 d{\varphi_0}_{\textrm{s}} \int^{2 \pi}_0 d \kappa_{\textrm{s}} \mathcal{F}(M_{\textrm{s}},\iota_{\textrm{s}},{\varphi_0}_{\textrm{s}},\kappa_{\textrm{s}}).
\end{equation}
Another useful metric to assess the closeness between waveforms is the
signal-to-noise (SNR)-weighted faithfulness \cite{Ossokine:2020kjp}
 \begin{widetext}
\begin{equation}
\overline{\mathcal{F}}_{\mathrm{SNR}}(M_\mathrm{s}) = \sqrt[3]{\frac{\int^{1}_{-1} d{(\cos \iota_{\textrm{s}})} \int_{0}^{2\pi} d\kappa_ {\mathrm{s}} \int_{0}^{2\pi} d{\varphi_0}_{\mathrm{s}} \ \mathcal{F}^{3}(M_{\textrm{s}},\iota_{\textrm{s}},{\varphi_0}_{\textrm{s}},\kappa_{\textrm{s}}) \ \mathrm{SNR}^3(\iota_{\textrm{s}},{\varphi_0}_{\textrm{s}},\kappa_{\textrm{s}})}{\int^{1}_{-1} d{(\cos \iota_{\textrm{s}})} \int_{0}^{2\pi} d\kappa_{\mathrm{s}} \int_{0}^{2\pi} d{\varphi_0}_{\mathrm{s}} \ \mathrm{SNR}^3(\iota_{\textrm{s}},{\varphi_0}_{\textrm{s}},\kappa_{\textrm{s}})}},
\label{eq:eq18}
\end{equation}
\end{widetext}
where the SNR is defined as
\begin{equation}
\mathrm{SNR}(\iota_{\textrm{s}},{\varphi_0}_{\textrm{s}},\theta_\textrm{s}, \phi_\textrm{s}, \kappa_{\textrm{s}},{D_{\mathrm{L}}}_{\mathrm{s}},\boldsymbol{\Theta}_\mathrm{s},{t_c}_\mathrm{s}) \equiv \sqrt{\left(h_{\mathrm{s}},h_{\mathrm{s}}\right)}.
\label{eq:eq19}
\end{equation}
In Eq. \eqref{eq:eq18} the weighting by the SNR takes into account the
dependence on the phase and effective polarization of the signal at a fixed
distance. Finally, we introduce the unfaithfulness or mismatch as
\begin{equation}
\overline{\mathcal{M}}=1-\overline{\mathcal{F}}.
\label{eq:eq20}
\end{equation}

\subsection{Comparison against quasi-circular numerical-relativity waveforms} \label{sec:NRquasicircular}

We begin by assessing the accuracy of the \texttt{SEOBNRv4EHM} waveform model  in the  zero
eccentricity limit, focusing on the unfaithfulness against the set of quasi-circular NR 
waveforms used to calibrate and validate the \texttt{SEOBNRv4HM} model. The public NR waveforms are available in the  \texttt{SXS} 
waveform catalog \cite{Boyle:2019kee} produced with the Spectral
Einstein code (\texttt{SpEC}) \cite{SpECwebsite}. The parameters of the public 
and non-public 141 NR waveforms are listed in the Appendix F of Ref.~\cite{Cotesta:2018fcv}.

In order to simplify our analysis, we restrict first to the $(2,|2|)$-modes
waveforms, and then include higher order modes. For the dominant $(2,|2|)$-modes
the faithfulness can be simplified with respect to Eq. \eqref{eq:eq16} as the
inclination angle of the signal is not usually considered due to the angular
dependence of the $_{-2}Y_{2 \pm2}$ harmonics, and the fact that $\kappa$, $\iota$ and $\varphi$ are degenerate
~\cite{PhysRevD.89.102003}. Therefore, the faithfulness function for the quasi-circular $(2,|2|)$-modes waveforms can be expressed as
\begin{equation}
\label{eq:eq21}
\mathcal{F}_{22}(M_{\textrm{s}},{\varphi_0}_{\textrm{s}}) =  \max_{t_c, {\varphi_0}_{t}} \left[\left . \frac{ \langle h_s|h_t \rangle}{\sqrt{  \langle h_s|h_s \rangle  \langle h_t|h_t \rangle}}\right \vert_{\substack{\boldsymbol{\Theta}_\mathrm{s}(t_{\mathrm{s}} = t_{0_\mathrm{s}}) = \boldsymbol{\Theta}_{t}(t_t = t_{0_\mathrm{t}})}} \right ].
\end{equation}
In practice, we remove the dependence of the faithfulness on the azimuthal angle
of the signal by evaluating Eq. \eqref{eq:eq21} in a grid of 8 values for
${\varphi_0}_{\textrm{s}} \in [0,2\pi]$, and averaging the result to obtain
$\overline{\mathcal{F}}_{22}$. The optimization over the coalescence time of the
signal is efficiently computed by applying an inverse Fourier Transform
\cite{Allen:2005fk} and we analytically optimize over the coalescence phase of
the template \cite{PhysRevD.89.102003}.
From Eq. \eqref{eq:eq21} one can define the
mismatch or unfaithfulness as,
\begin{equation}
\overline{\mathcal{M}}_{22}=1-\overline{\mathcal{F}}_{22}.
\label{eq:eq22}
\end{equation}
The condition $\boldsymbol{\Theta}_\mathrm{s}(t_{\mathrm{s}} =
t_{0_\mathrm{s}}) = \boldsymbol{\Theta}_{t}(t_t = t_{0_\mathrm{t}}) $
in Eq. \eqref{eq:eq21} enforces that the intrinsic parameters of both
the template and the signal are the same at the start of the waveform
$t=t_0$. This implies that the component masses $m_{1,2}$, or
equivalently the mass ratio $q=m_1/m_2\geq 1$ and the total mass
$M=m_1+m_2$, and the dimensionless spins $\chi_{1,2}$ of the
signal and the template are identical at $t_0$~\cite{Ossokine:2020kjp}.

\begin{figure}[htbp!]
\begin{center}
\includegraphics[width=1.\columnwidth]{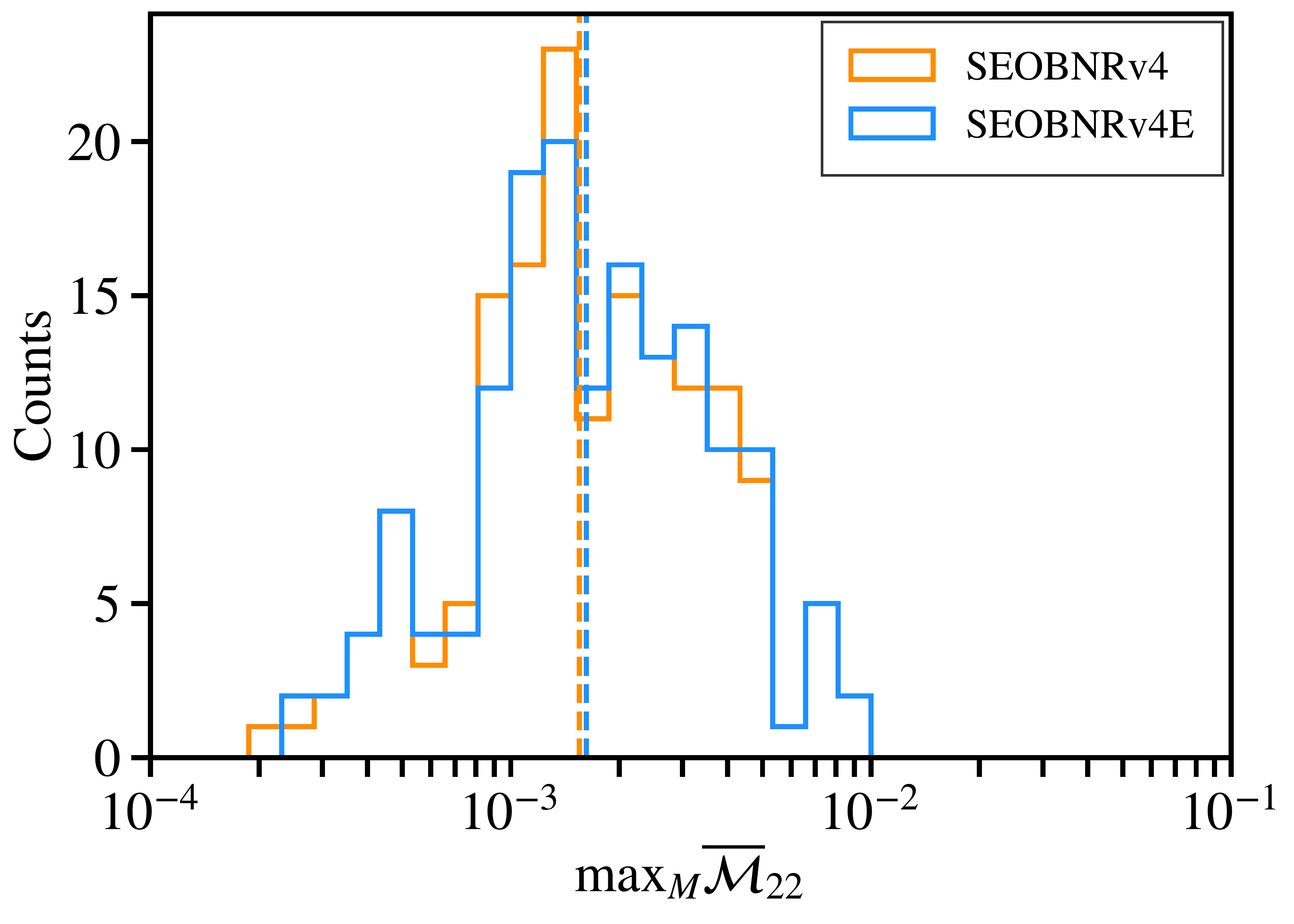}
\caption{Distribution of the maximum unfaithfulness of
\texttt{SEOBNRv4} (orange) \cite{Bohe:2016gbl} and the new \texttt{SEOBNRv4E} (blue)
against the public 141 quasi-circular NR simulations of Ref. \cite{Bohe:2016gbl}. The
total mass range considered is $20 M_\odot \leq M \leq 200 M_\odot$. The
calculations are done with  the Advanced LIGO's zero-detuned high-power design
sensitivity curve \cite{Barsotti:2018}. The vertical dashed orange (blue)
lines correspond to the median values of the \texttt{SEOBNRv4}
(\texttt{SEOBNRv4E}) distributions. }
\label{fig:QC22mismatches}
\end{center}
\end{figure}

For the calculation of the unfaithfulness we consider a total-mass range of  $20
M_\odot \leq M \leq 200 M_\odot$. We show in Fig. \ref{fig:QC22mismatches} 
the $22-$mode unfaithfulness maximized over the total-mass range for the 
\texttt{SEOBNRv4} and \texttt{SEOBNRv4E} models. We remind that the \texttt{SEOBNRv4} model 
was calibrated requiring an unfaithfulness for the $22$-mode against the 141 NR waveforms 
of at most of $1\%$. It is interesting to note that the \texttt{SEOBNRv4E} eccentric 
model achieves a similar accuracy, with the median of the distribution 
slightly larger than the one of the \texttt{SEOBNRv4} model. This is due to the fact that the 
\texttt{SEOBNRv4E} model contains eccentric corrections, which were not available 
when calibrating the \texttt{SEOBNRv4} model to NR in the quasi-circular limit.

\begin{figure}[htbp!]
\begin{center}
\includegraphics[width=1.\columnwidth]{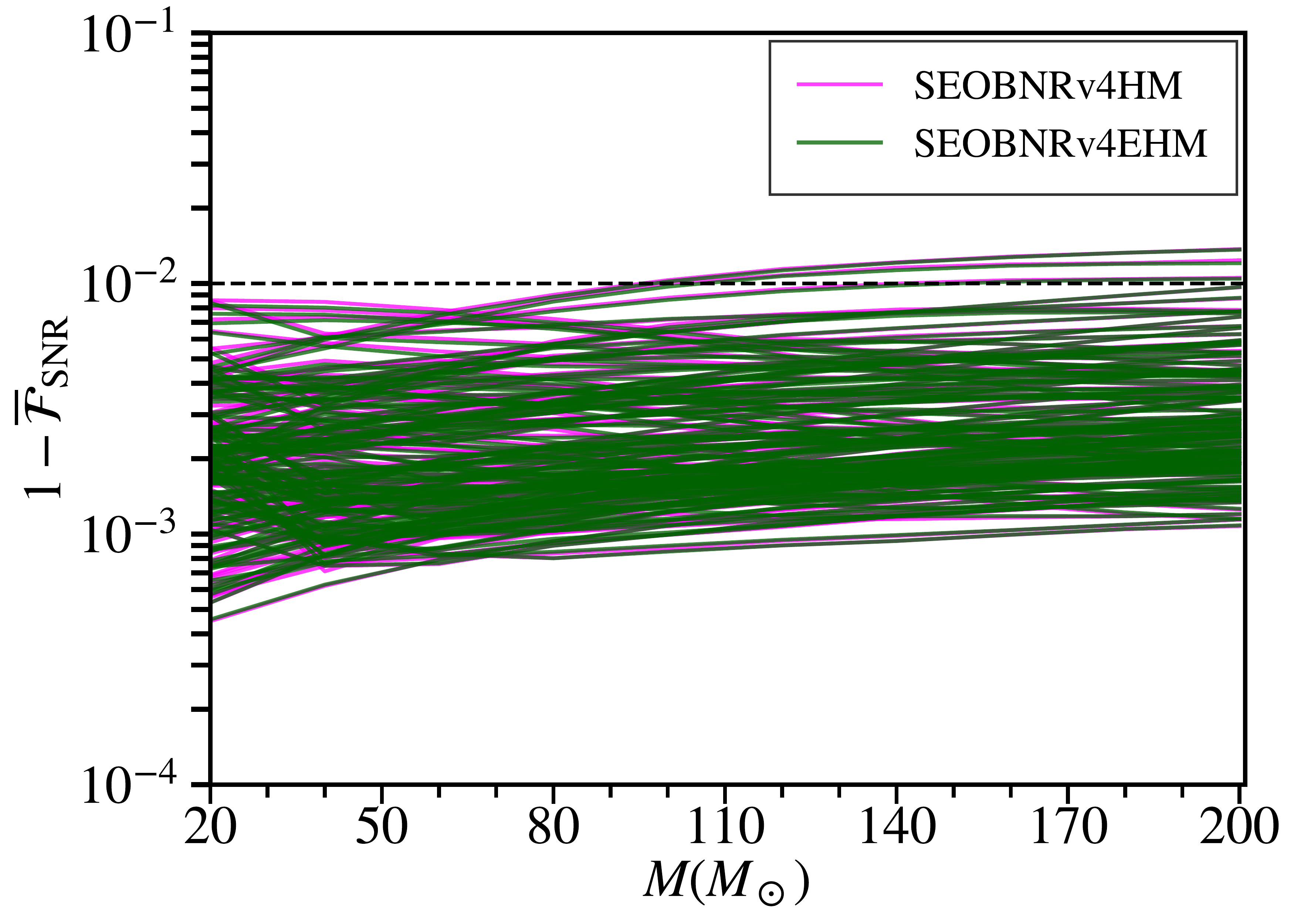}
\caption{SNR-weighted unfaithfulness, as defined in Eq. \eqref{eq:eq20}, as a function of the total mass, in the range $20 M_\odot \leq M \leq 200 M_\odot$, between the \texttt{SEOBNRv4HM} (pink) and \texttt{SEOBNRv4EHM} (green) models and the SXS quasi-circular NR waveforms used in Ref. \cite{Cotesta:2018fcv}. The calculations are done with the Advanced LIGO's zero-detuned high-power design-sensitivity curve \cite{Barsotti:2018}, which is an estimate for the upcoming O4 run. The horizontal black dashed line 
indicates the $1\%$ unfaithfulness value.}
\label{fig:QCHMmismatches}
\end{center}
\end{figure}

The higher order multipoles included in \texttt{SEOBNRv4EHM} are the same as in
\texttt{SEOBNRv4HM} (i.e., $(l,|m|)=\{(2,1),(3,3),(4,4),(5,5)\}$). When computing the unfaithfulness
of the models with higher modes against NR, we include in the NR waveforms all the modes with $l\leq 5$ as done
in Ref.~\cite{Cotesta:2018fcv}. To ease the visualization of the results, we compute the 
SNR-weighted mismatches defined in Eq.~\eqref{eq:eq18}, and average over the signal inclination, azimuthal and
effective polarization angles. In  practice, the average is performed over three different inclination angles of
the signal $\iota_s =\{0, \pi/3,\pi/2\}$, and for each inclination angle we make
a grid of $8\times 8$ for $\kappa_s,\varphi_{0s} \in [0,2 \pi]$.  In Fig.~\ref{fig:QCHMmismatches}, we show 
the SNR-weighted mismatches for the \texttt{SEOBNRv4HM}  and  \texttt{SEOBNRv4EHM} models against the quasi-circular 
NR waveforms at our disposal. Again, we note that the mismatches of the \texttt{SEOBNRv4EHM} model 
are very similar to the ones of the \texttt{SEOBNRv4HM} model.  There are a few cases at high total masses for 
which both models have unfaithfulness above $1\%$, but not larger than $1.5 \%$, as reported also in Ref.~\cite{Cotesta:2018fcv}. 
This indicates that the higher-order modes in the eccentric model have a comparable accuracy to the ones of the underlying quasi-circular
model in the zero eccentricity limit. 

\subsection{Comparison against eccentric numerical-relativity waveforms}
\label{sec:NReccentric} 

The calculation of the unfaithfulness assumes that the
intrinsic parameters of both template and signal are identical at the start of
the evolution, that  is we use the condition $\boldsymbol{\Theta}_\mathrm{s}(t_{\mathrm{s}}
= t_{0_\mathrm{s}}) = \boldsymbol{\Theta}_{t}(t_t = t_{0_\mathrm{t}}) $ in Eq.
\eqref{eq:eq16}. In the eccentric case, this would imply that the mass ratio,
$q$, total mass, $M$, dimensionless spins, $\chi_{1,2}$, eccentricity, $e$
and relativistic anomaly, $\zeta$, of both the signal and the template are the same at the
start of the waveform. While the spins and mass
parameters are uniquely fixed in the non-precessing spinning case, the
eccentricity and relativistic anomaly cannot be uniquely identified
with respect to the NR waveforms. Consequently, when comparing a waveform model
against eccentric NR waveforms~\footnote{Except in the case of the eccentric NR surrogate
model \cite{Islam:2021mha}, which is constructed with the same definitions of
eccentricity and mean anomaly used to measure these parameters from NR
waveforms.}, an optimization over the initial eccentricity and relativistic anomaly has
to be performed to take into account the different definition of eccentricity
between the model and the NR waveforms.

In the case of the \texttt{SEOBNRv4EHM} model, to reduce the
dimensionality of the parameter space, we use the initial conditions for
eccentric orbits starting at periastron ($\zeta=0$), and we compute the eccentric
EOB waveforms by specifying the initial eccentricity, $e_0$, and
starting frequency, $\omega_0$. Thus, when computing the faithfulness
of the model against eccentric NR waveforms we have to maximize over
$e_0$ and $\omega_0$. Furthermore, here we compute the unfaithfulness using
also another publicly available eccentric EOB model, \texttt{TEOBResumSE}~\footnote{In this work, we use the
  \texttt{eccentric} branch of the public bitbucket repository
  \url{https://bitbucket.org/eob_ihes/teobresums} with the git hash
  \texttt{39e6d7723dacb23220ff5372e29756e5f94cb004}, which is the latest at the time 
of this publication.}
\cite{Chiaramello:2020ehz,Nagar:2021gss,Nagar:2021xnh, Placidi:2021rkh}, for which we
also specify initial conditions at periastron, and optimize over the
initial eccentricity and starting frequency of the model. We note that
although the \texttt{TEOBResumSE} can include higher-order modes, we use here
only the $(2,|2|)$-modes, as we have found that in the presence of
eccentricity some of the higher-order modes develop unphysical
behaviors close to merger and ringdown. These features are likely due 
the treatment of the NQC corrections in the eccentric case, as already reported in 
Ref.~\cite{Nagar:2021xnh}.

First, we focus on the $(2,|2|)$-modes only waveforms. We define the eccentric
faithfulness function as follows,
\begin{equation}
\label{eq:eq23}
\mathcal{F}^{\text{ecc}}_{22}(M_{\textrm{s}},{\varphi_0}_{\textrm{s}}) =  \max_{t_{0_t}, {\varphi_0}_{t},e_0,\omega_0} \left[\left . \frac{ \langle h_s|h_t \rangle}{\sqrt{  \langle h_s|h_s \rangle  \langle h_t|h_t \rangle}}\right \vert_{\substack{\boldsymbol{\Theta}_\mathrm{s}(t_{\mathrm{s}} = t_{0_\mathrm{s}}) = \boldsymbol{\Theta}_{t}(t_t = t_{0_\mathrm{t}})}} \right ].
\end{equation}
For completeness, we introduce here also the unfaithfulness function as
\begin{equation}
\overline{\mathcal{M}}^{\text{ecc}}_{22}=1-\overline{\mathcal{F}}^{\text{ecc}}_{22}.
\label{eq:eq24}
\end{equation}

\begin{figure}[htbp!]
\begin{center}
\includegraphics[width=1.\columnwidth]{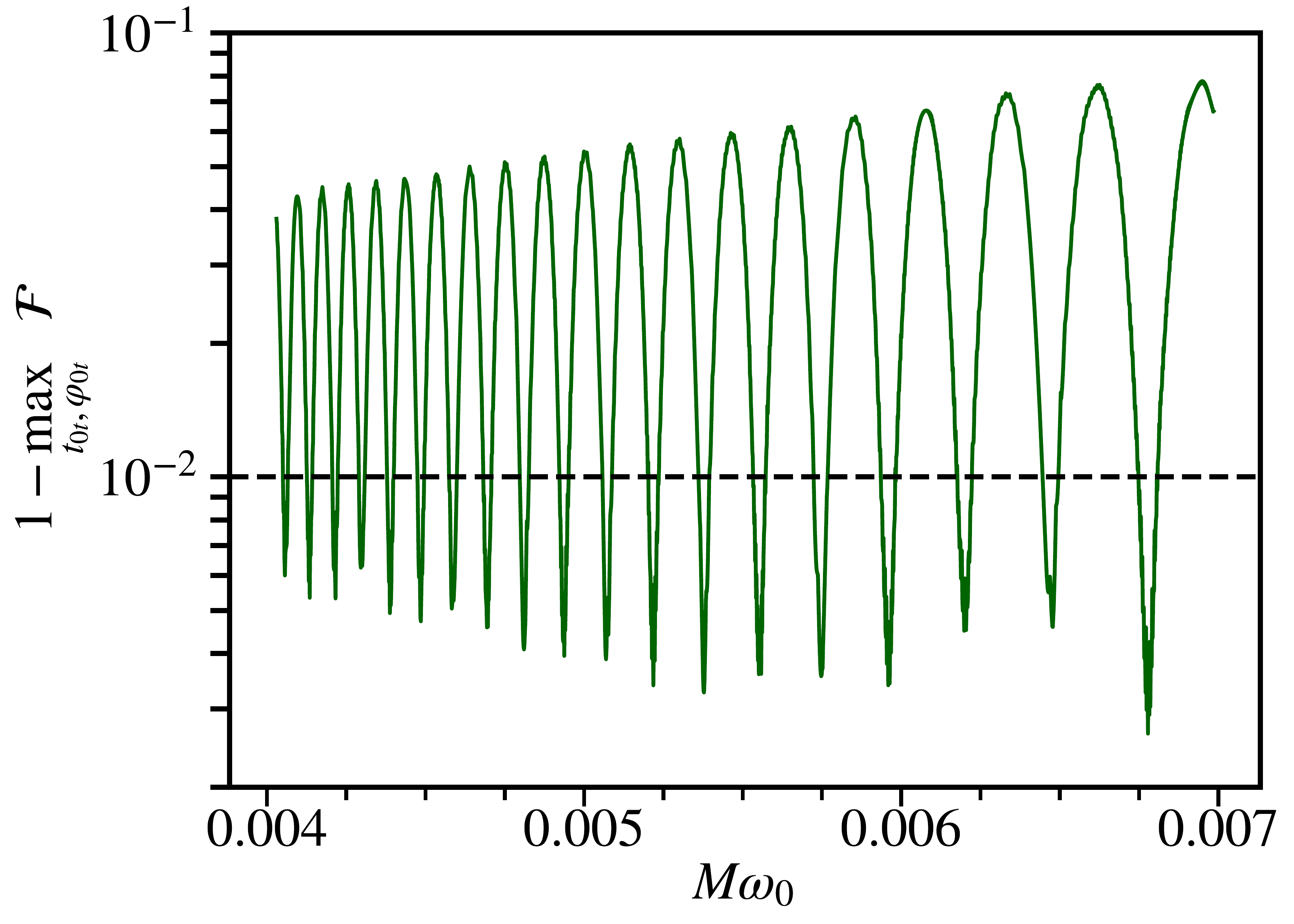}
\caption{Unfaithfulness of the \texttt{SEOBNRv4E} waveform model against the \texttt{SXS:BBH:1355} waveform at a fixed initial eccentricity, $e_0=0.09$, as a function of the starting frequency of the model $M \omega_0$. The unfaithfulness is computed at a fixed total mass of $20 M_\odot$, maximizing over the coalescence time, $t_{0_t}$, and azimuthal angle, $\varphi_{0_t}$,  of the model.}
\label{fig:MatchOscillatory}
\end{center}
\end{figure}

From Eq. \eqref{eq:eq23} one observes that in the eccentric case two additional
numerical optimizations have to be performed, as compared to the quasi-circular
case (see Eq. \eqref{eq:eq21}). The main difficulty of estimating such optimal
values arises from the fact that the  two additional optimizations cannot be
easily performed with standard optimization algorithms, as the
unfaithfulness has a highly oscillatory behavior as a function of these
parameters. This can be observed in Fig. \ref{fig:MatchOscillatory} where we show 
the unfaithfulness of the \texttt{SEOBNRv4E} waveform against the \texttt{SXS:BBH:1355}
waveform as function of the starting frequency. In Fig.
\ref{fig:MatchOscillatory}, the mismatch is computed for a total mass of
$20M_\odot$ and at fixed initial eccentricity $e_0=0.09$ for the \texttt{SEOBNRv4E} model. The
high number of local maxima and minima in the unfaithfulness function makes standard
optimization algorithms quite inefficient, and it increases substantially the
computational cost of such procedure.

In order to overcome this problem, different eccentric EOB models use different
approaches to estimate the optimal values for $e_0$ and $\omega_0$. In the case
of the \texttt{SEOBNREHM} model \cite{Cao:2017ndf, Liu:2019jpg,Liu:2021pkr} (not to be 
confused with our \texttt{SEOBNRv4EHM} model here), the starting frequency is set
to the frequency when the eccentricity is estimated from the NR waveforms, and
then the initial eccentricity is varied to get the best match against the NR
waveforms. While for the \texttt{TEOBResumSE} model, the eccentricity and starting frequency are
varied manually to get the lowest unfaithfulness against NR \cite{Nagar:2021gss}.

Here, we develop an automatic procedure to perform the two optimizations over
$e_0$ and $\omega_0$. The procedure is as follows:
\begin{itemize}
\item[1)] Fix the total mass to the lower bound of the total mass range used, that is $20M_\odot$.
In this way, we ensure that more inspiral part of the NR waveform is in the frequency band of the mismatch calculation.
\item[2)] Create a grid in eccentricity of $N_e$ values around the value of the
eccentricity as measured from the NR orbital frequency using Eq.
\eqref{eq:eq13}, $e^{\text{NR}}_{\omega_{\text{orb}}}$, such that $e_0 \in
[e^{\text{NR}}_{\omega_{\text{orb}}}- \delta e,
e^{\text{NR}}_{\omega_{\text{orb}}}+\delta e]$. The value of $\delta e $
determines the eccentricity interval. In the case
$e^{\text{NR}}_{\omega_{\text{orb}}}- \delta e<0$, we take the lower bound to be
zero.
\item[3)] For each value of the eccentricity, generate a grid of $N_\omega$ values of
starting frequency. The upper bound is determined by the frequency at which the
EOB waveform equals the length $l$ of the eccentric NR waveform,
$\omega^{l_{\text{NR}}=l_{\text{EOB}}}_0$, while the lower bound is determined by a chosen
$\delta \omega$. Thus, the frequency grid is  $\omega_0 \in
[\omega^{l_{\text{NR}}=l_{\text{EOB}}}_0 - \delta \omega, \omega^{l_{\text{NR}}=l_{\text{EOB}}}_0]$.
\item[4)] For each point in the grid, compute the unfaithfulness optimizing over
the time shift and phase offset of the template as in Eq. \eqref{eq:eq21}.
\item[5)] Store the values $(e^{\text{opt}}_0,\omega^{\text{opt}}_0)$ which
provide the lowest unfaithfulness.
\item[6)] In order to reduce the computational cost, we use the optimal values,
$(e^{\text{opt}}_0,\omega^{\text{opt}}_0)$  at $20 M_\odot$ for the whole mass
range.
\end{itemize}

For the results in this paper we choose $N_e=200$, $N_f=500$, an eccentricity interval, $\delta e =0.1$, and a starting frequency interval of $\delta f =10$Hz at $20M_\odot$, which translates into
$\delta \omega = 0.006$.

The above optimization procedure is tested by computing the unfaithfulness against  the
eccentric NR waveforms publicly available in the SXS catalog
\cite{Hinder:2017sxy,Boyle:2019kee}. In Table \ref{tab:tabNR} we summarize the
main properties of the NR simulations used in this work,  the  optimal values of
initial eccentricity and starting frequency of the \texttt{SEOBNRv4E} and \texttt{TEOBResumSE} models, and the maximum value over total mass range of the unfaithfulness of the models against
the NR simulations. We note that there is a particular simulation \texttt{SXS:BBH:1169} for which the optimization procedure leads to zero initial eccentricity for \texttt{TEOBResumSE}. This is a NR simulation with very low eccentricity for which the quasi-circular waveform has already a very low mismatch of $0.78 \%$. The reported value of the unfaithfulness of \texttt{TEOBResumSE} for this case in Table \ref{tab:tabNR} is slightly lower than that reported in Refs.~\cite{Nagar:2021gss,Nagar:2021xnh}. We have checked modifications of the grid parameters in the eccentricity and starting frequency grids for the optimization procedure, particularly increasing the resolution up to $N_e=300$ and $N_f=1000$, but this optimal eccentricity value of zero still remains unchanged. While it is possible that further increasing the resolution could lead to a slightly lower unfaithfulness, it also significantly increases the computational cost and therefore we opt to use the already calculated value.  We also  note that for this particular case (which has low eccentricity and high spins) the optimization procedure may  be affected by the slight discontinuity of the \texttt{TEOBResumSE} model for small eccentricities as already noted in Ref.~\cite{OShea:2021ugg}.

\begin{figure}[h]
\begin{center}
\includegraphics[width=1.\columnwidth]{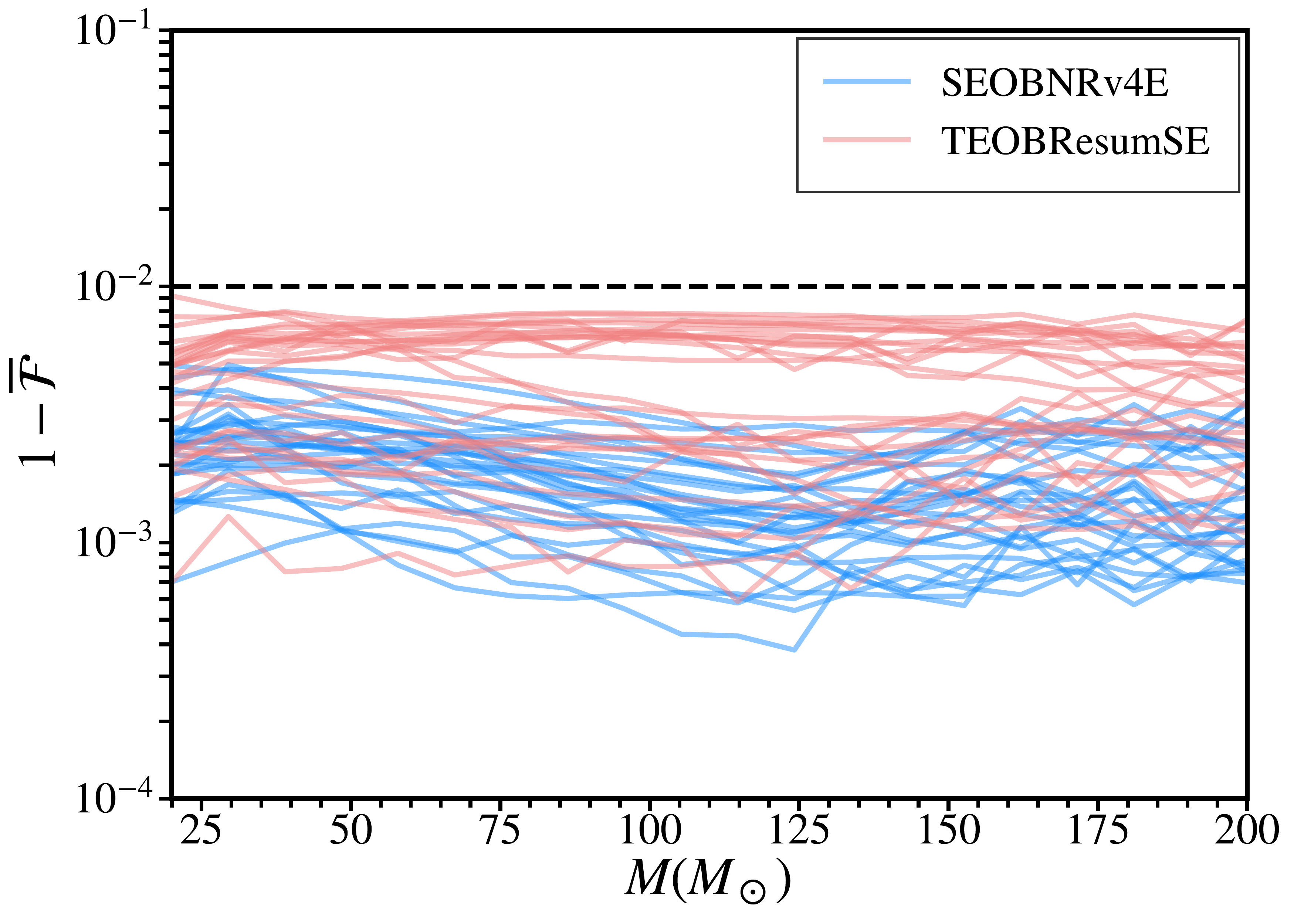}
\includegraphics[width=1.\columnwidth]{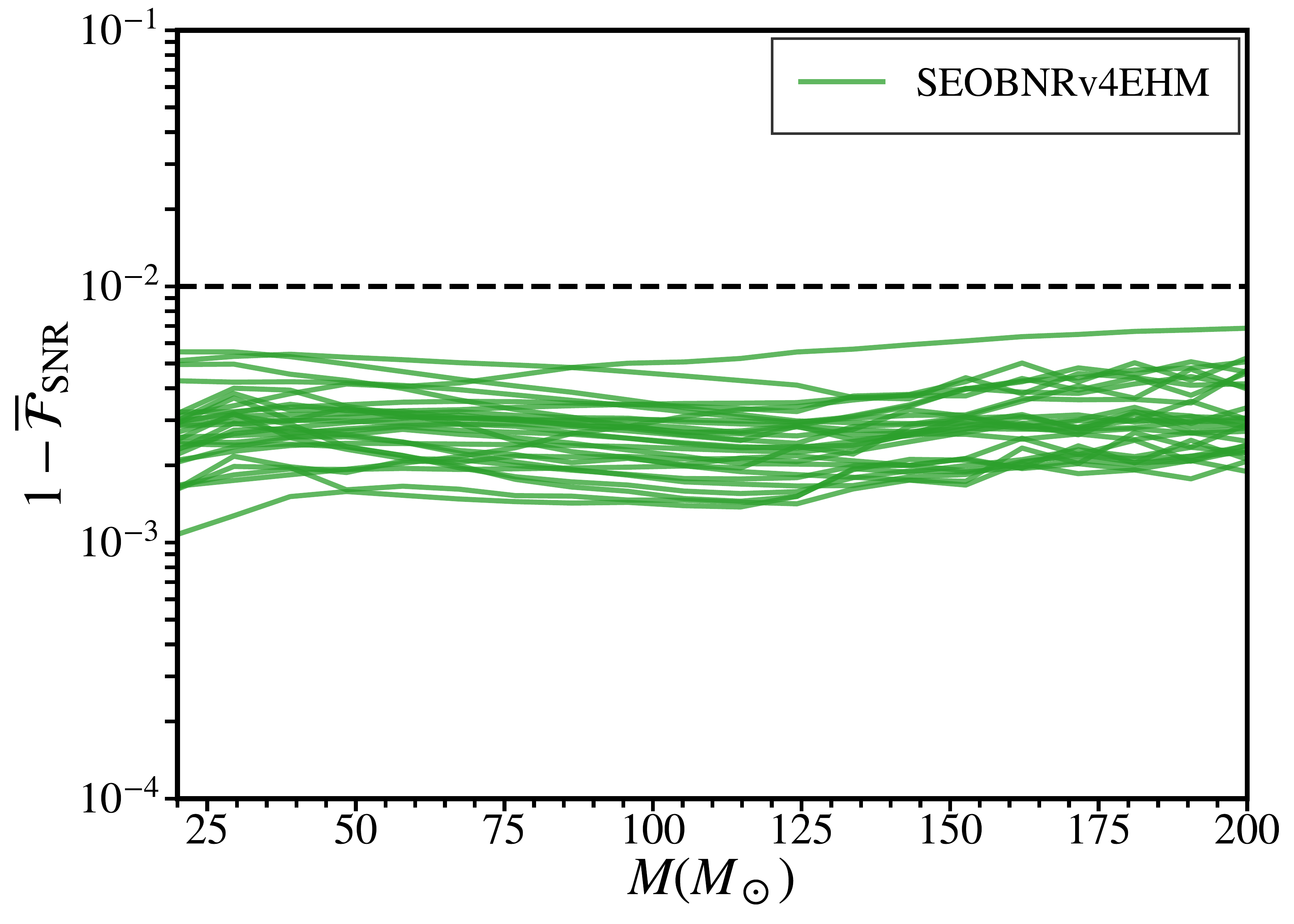}
\caption{Upper panel: Unfaithfulness of the \texttt{SEOBNRv4E} and \texttt{TEOBResumSE} models against the 28 eccentric public SXS simulations listed in Table \ref{tab:tabNR}. The calculations are performed optimizing over the initial eccentricity and starting frequency at periastron. Lower panel: Unfaithfulness of \texttt{SEOBNRv4EHM} model against the same NR waveforms as in the upper panel, but including all modes with $l \leq 5$ in the NR waveforms. The calculations are performed using the optimal values of eccentricity and starting frequency obtained from the unfaithfulness computed with \texttt{SEOBNRv4E}. The horizontal dashed lines in both panels indicate the $1\%$ value of unfaithfulness.}
\label{fig:EccMismatches}
\end{center}
\end{figure}

The unfaithfulness of the \texttt{SEOBNRv4E} and
\texttt{TEOBResumSE} models against the dataset of eccentric NR
waveforms described in Table \ref{tab:tabNR} are shown in the upper
panel of Fig. \ref{fig:EccMismatches}. We note that the unfaithfulness
curves are always below $1\%$ for the whole dataset and total mass
ranges. These results also indicate that the approximation in step 6)
of using the same optimal values
for $(e^{\text{opt}}_0,\omega^{\text{opt}}_0)$ for the whole total
mass range is reasonable, as the unfaithfulness does not significantly
increase with the total mass range. The bulk of the unfaithfulness
curves for the \texttt{SEOBNRv4E} model is below the ones of the
\texttt{TEOBResumSE} model for the NR dataset considered here. We note
that the values of the unfaithfulness for the \texttt{TEOBResumSE}
model reported here are similar to the ones in
Ref.~\cite{Nagar:2021gss}. However, in recent
publications~\cite{Nagar:2021xnh,Placidi:2021rkh}, lower unfaithfulnesses are reported
for the \texttt{TEOBResumSE} model, driven by recalibrating it to
quasi-circular NR waveforms, and by better computing the Fourier
transform of the time-domain waveforms, as remarked in Refs.~\cite{Nagar:2021xnh,Placidi:2021rkh}.
(This improved model is not
public.) As a consequence of those improvements, the main bulk of the
unfaithfulness curves is closer to $10^{-3}$ values, and thus, at a 
similar level as the \texttt{SEOBNRv4E} model in
Fig.~\ref{fig:EccMismatches}.  We remark that in order to better
assess the accuracy of both models, comparisons against larger
datasets of eccentric NR simulations are required. Eventually,
Bayesian inference analyses will be needed to assess biases in the
recovered parameters.

The calculation of the unfaithfulness including higher order modes
requires three numerical optimizations (initial eccentricity, starting
frequency and azimuthal angle of the template) and an analytical optimization
over the effective polarization angle for each single point in the sky of the
signal. Consequently, computing SNR-weighted unfaithfulness averaged over the
sky-positions, orientations and inclinations of the signal becomes
computationally prohibitive. In order to reduce the computational cost, we
assume that the optimal values for the initial eccentricity and starting frequency
of the \texttt{SEOBNRv4EHM} model are the same as the ones obtained for the 
\texttt{SEOBNRv4E} model, $(e^{\text{opt}}_0,\omega^{\text{opt}}_0)$, and  we compute the unfaithfulness
as in the quasi-circular case, numerically optimizing over the azimuthal angle of
the template, and analytically over the effective polarization angle of the
template. 

\begin{figure*}[htbp!]
\begin{center}
\includegraphics[width=2\columnwidth]{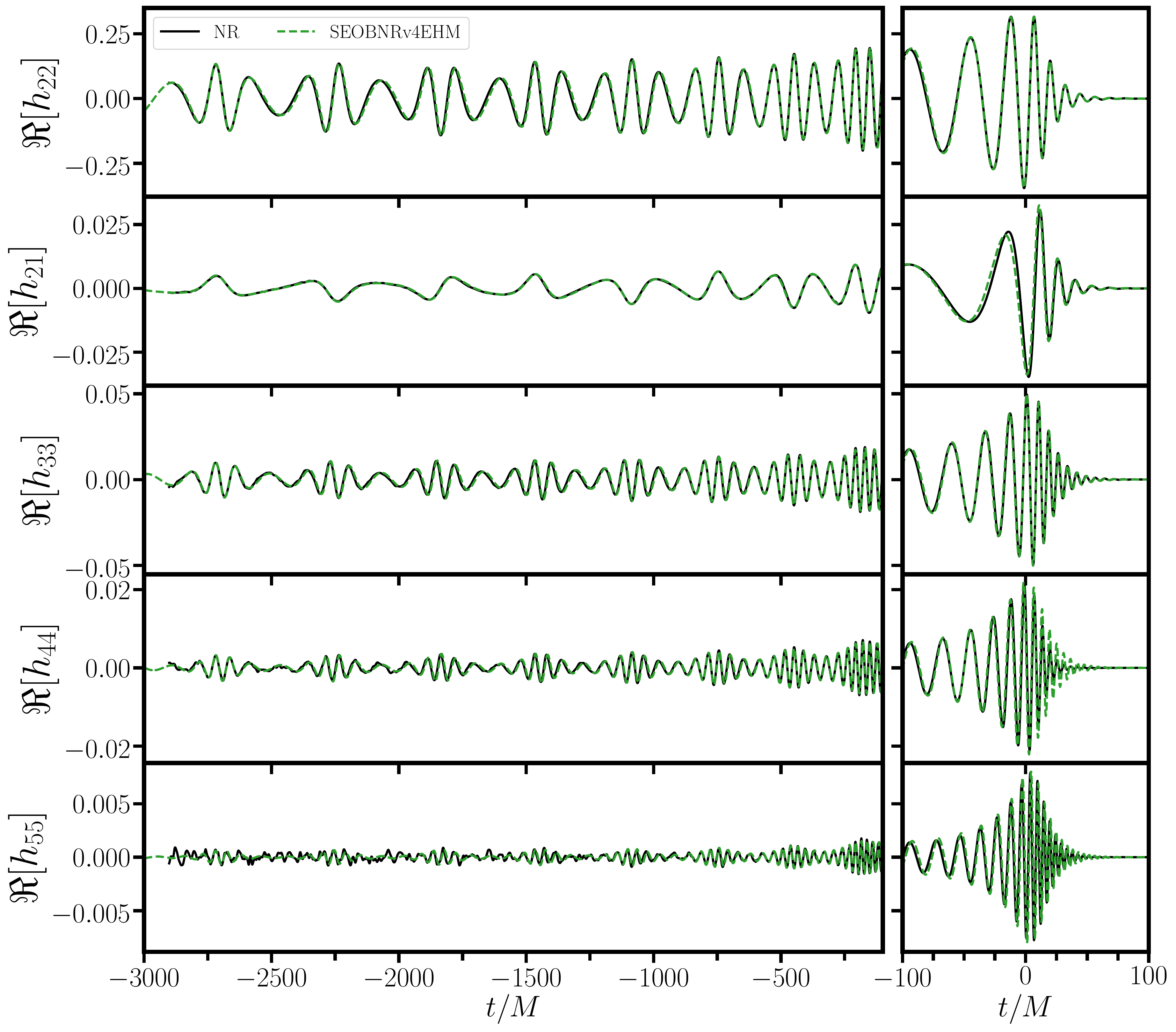}
\caption{From top to bottom, real part of the (2,2),(2,1), (3,3), (4,4) and (5,5)
modes in the time domain. The black curve corresponds to the NR simulation \texttt{SXS:BBH:1369}, which has mass ratio $q = 2$, zero spins, and eccentricity $e_{\omega_\text{orb,p}}=0.257$, while the green curve corresponds to the \texttt{SEOBNRv4EHM} model for the values of eccentricity
and starting frequency that lead to the lowest unfaithfulness for the
(2,2) mode.}
\label{fig:higherModes}
\end{center}
\end{figure*}

We apply this approximation and compute the unfaithfulness between the
\texttt{SEOBNRv4EHM} waveforms and the eccentric NR waveforms, which
include all the modes with $l \leq 5$. We show the results in the
lower panel of Fig.~\ref{fig:EccMismatches}. We can observe that the
curves of the SNR-weighted unfaithfulness for the multipolar model are
always below $1 \%$. This indicates that the approximation of using
the optimal values of $(e^{\text{opt}}_0,\omega^{\text{opt}}_0)$
obtained from the unfaithfulness of the \texttt{SEOBNRv4E} model is a
good approximation.  When comparing the unfaithfulness of the \texttt{SEOBNRv4EHM} model 
against the {\tt SEOB} model developed in Ref. \cite{Liu:2021pkr} (\texttt{SEOBNREHM}), we find 
that the unfaithfulness of \texttt{SEOBNRv4EHM} are always smaller than $1\%$, which is not the case for the \texttt{SEOBNREHM} model, which presents some cases with unfaithfulness as large as $2 \%$~\cite{Liu:2021pkr}. We also note that the unfaithfulness for the
\texttt{SEOBNRv4EHM} model is overall larger than that for the
\texttt{SEOBNRv4E} model, this may indicate that the higher-order
modes in the multipolar model are not as accurately modeled as the
dominant $(2,2)$ mode. However, we remark that the procedure to
compute the unfaithfulness for the model with higher-order modes is
suboptimal as the values of $(e^{\text{opt}}_0,\omega^{\text{opt}}_0)$
are obtained from the \texttt{SEOBNRv4E} model, thus the
unfaithfulness results for \texttt{SEOBNRv4EHM} are a conservative
estimate. Furthermore, some higher-order modes in the dataset of the
eccentric NR waveforms are affected by numerical noise, which may also
affect the unfaithfulness values. This can also be seen in
Fig.~\ref{fig:higherModes} where the different multipoles of the
\texttt{SEOBNRv4EHM} waveform model for the optimal values of
$(e^{\text{opt}}_0,\omega^{\text{opt}}_0)$, and of the NR waveform
\texttt{SXS:BBH:1364} are shown. We note that the higher-order modes of the model 
have very good agreement with respect to NR, and that the early inspiral of the NR
(5,5)-mode is dominated by numerical noise. 
  Thus, larger datasets of more accurate
eccentric NR waveforms are required in order to better assess and
improve the accuracy of multipolar eccentric EOB waveform models.

\subsection{Robustness of the model across parameter space}
\label{sec:Robustness}

Having assessed the accuracy of the model against the NR waveforms at our disposal, we now explore the region of validity of the 
\texttt{SEOBNRv4EHM} waveform model and identify the regions of parameter space where the model can be robustly generated.

One important property of a waveform model is its smoothness under small
  perturbations of the intrinsic parameters. In order to test this
  property, we compute the unfaithfulness between the 
  \texttt{SEOBNRv4E} waveforms perturbing the
  eccentricity parameter by $\delta e = 10^{-7}$. We perform this test
  using the \texttt{pycbc\_faithsim} function in the \texttt{PyCBC}
  software \cite{alex_nitz_2020_3630601} and employ for $10^6$
  waveforms randomly distributed in the following parameter space:
  $\chi_{1,2} \in [-0.99,0.99]$, $q \in [1,20]$, $M\in
  [10,100]M_\odot$, $e_0\in [0,0.3]$. We choose $19$Hz for the
  starting frequency of the waveforms, and $20$Hz for the overlap
  calculations. We find that only a few cases have unfaithfulness
  above $10^{-8}$, with the maximum mismatch being $0.3 \%$ and the
  median mismatch being $0$, indicating that the waveform model
  behaves smoothly under changes of the eccentricity parameter.

\begin{figure*}[!]
\begin{center}
\includegraphics[width=2\columnwidth]{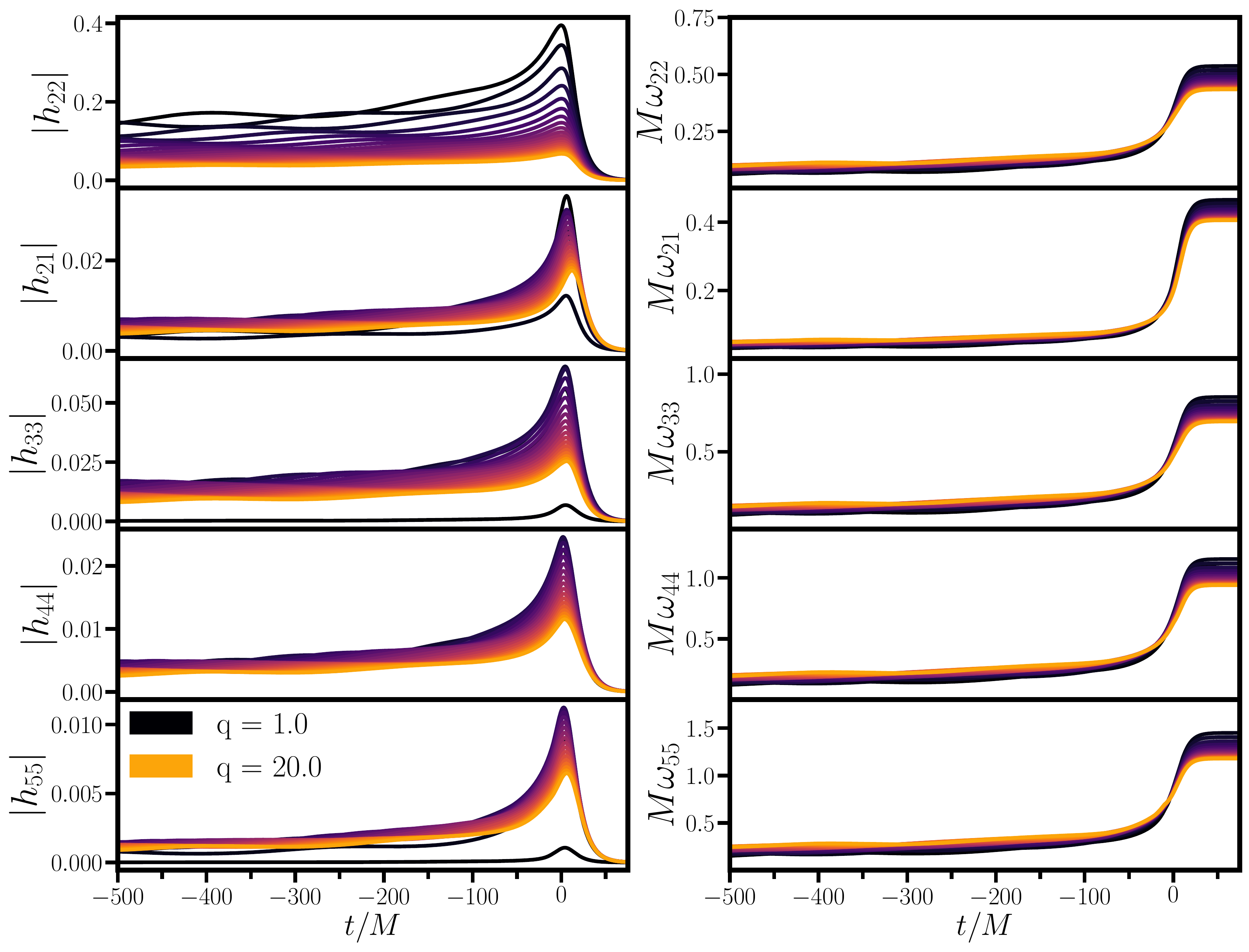}
\caption{From top to bottom amplitudes (left panels) and frequencies (right panels) of the $(2,2)$, $(2,1)$, $(3,3)$, $(4,4)$ and $(5,5)$ modes of \texttt{SEOBNRv4EHM} versus time, aligned at merger, for a configuration with initial eccentricity  $e_0=0.25$, spins $(\chi_1,\chi_2)= (0.5,-0.75)$, starting frequency $20$Hz and total mass $60 M_\odot$, for a mass ratio range $q\in[1,20]$.}
\label{fig:SmoothVariation}
\end{center}
\end{figure*}

As an example, we show in Fig.~\ref{fig:SmoothVariation} the
  amplitude and frequency of the multipoles in the
  \texttt{SEOBNRv4EHM} model (i.e.,
  $(l,m)=\{(2,2),(2,1),(3,3),(4,4),(5,5)\}$ modes), as function of time and 
aligned at the merger time,  for different values of 
the mass ratio, for a configuration
  with initial eccentricity $e_0=0.25$, spins $(\chi_1,\chi_2)=
  (0.5,-0.75)$, starting frequency $20$Hz and with total mass $60
  M_\odot$. As can be seen, the modes of the \texttt{SEOBNRv4EHM} model have a
  smooth behaviour in amplitude and frequency under variation of the
  mass ratio $q = 1 \mbox{--}20$. We note that, during the inspiral, in the equal-mass
  case, the amplitude of the odd-m modes (except the $(2,1)$ mode)
  is very small compared to the unequal-mass
  configurations. The same behavior is present in the amplitudes of
  the quasi-circular orbit \texttt{SEOBNRv4HM} model. As discussed in 
Ref.~\cite{Cotesta:2018fcv} (see Fig.~2 and text around), this is due to the fact that this
  binary configuration has a relatively large asymmetric-spin parameter $\chi_A = (\chi_1-\chi_2)/2=0.5$, for
  which, in the equal-mass (or nearly equal-mass) case 
the odd-m modes (except the $(2,1)$-mode) have very small amplitude during the inspiral, as predicted by PN theory.

We find that the  orbit averaging procedure  that we apply to the NQC function works 
quite well from large negative spins to mild positive spins, but binary's configurations 
with large-positive spins and small initial separation (or large dimensionless orbital 
frequency) can challenge this procedure due to the last periastron passage occurring 
very close to merger (see the Appendix \ref{sec:AppendixA}). This can cause oscillations in the dynamical quantities in the late 
inspiral. In order to test the smoothness of the waveform model in the large-spin region, 
we compute the unfaithfulness between two waveforms varying the spins in the region
$\chi_{1,2}\in[0.8,0.99]$ for 100 mass ratios $q\in[1,50]$. For each
mass ratio, we compute the unfaithfulness between a waveform with 
$\chi_1 = \chi_2=0.8$, initial eccentricity $0.3$ at starting frequency of 20Hz and
total mass $100M_\odot$ and waveforms with the same parameters but varying both 
$\chi_{1,2} \in [0.8,0.99]$. This choice of total mass, starting frequency
and eccentricity implies smaller initial separations of $r/M \sim 11$,
and thus corresponds to a challenging case for the quasi-circular
assumption of the merger-ringdown signal. The results from such
a test show an oscillatory unfaithfulness surface across parameter
space without sharp features. We also observe that for $\chi_{1,2}
\gtrsim 0.9 \mbox{--} 0.95$ the frequency of the (2,2)-mode can have small
spurious oscillations, thus, the model should be used with
caution in this region of parameter space. Nevertheless, the
model does not show prominent features in the waveform, and therefore,
we recommend that it is used up to spins $0.99$, eccentricity $0.3$ and 
initial frequency up to $20$ Hz. We plan to improve the model in the 
transition from plunge to merger for large spins, as soon as  
we will have access to NR eccentric waveforms with large spins.

\begin{figure}[!]
\begin{center}
\includegraphics[width=1\columnwidth]{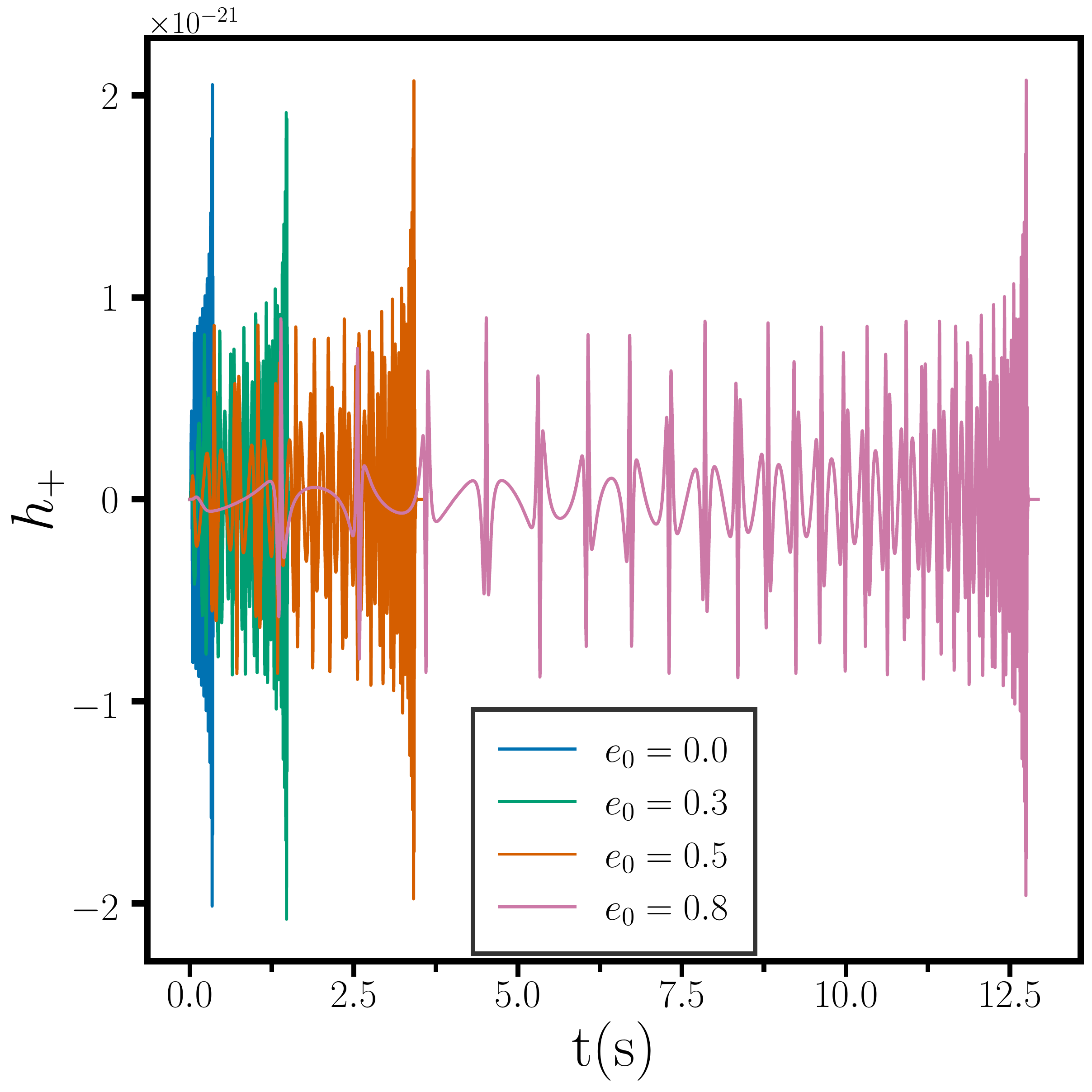}
\caption{Plus gravitational polarization versus time for a $q=2$
  non-spinning configuration computed with \texttt{SEOBNRv4EHM} for
  four different initial eccentricities $e_0=[0,0.3,0.5,0.8]$,
  and total mass $M=100M_\odot$. All the configurations have a starting frequency of $20$Hz at periastron. We note that for $e_0=0.8$ the waveform reproduces the burst-like features
  produced at the periastron passages.}
\label{fig:polarizationHighEccentricity}
\end{center}
\end{figure}

We note that whereas we have probed the validity of the model through comparisons 
to the public SXS eccentric waveforms and internal consistency tests mostly 
for mass ratios $q\in[1,20]$ and eccentricities $e\in[0,0.3]$ at 20Hz, we can
also generate \texttt{SEOBNRv4EHM} waveforms at higher eccentricities and mass ratios, 
as illustrated in Fig.~\ref{fig:polarizationHighEccentricity}, where we show 
the plus polarization $h_+$ for a non-spinning BBH with mass-ratio 2 
and different initial eccentricities $e_0=[0,0.3,0.5,0.8]$. These configurations are produced with a starting frequency of 20Hz defined at periastron, so that, at that time, there are no frequencies in the inspiral higher than the starting frequency. In fact, 
  we have checked that the model can be robustly generated at higher eccentricities by producing a large set of  ($10^6$) waveforms randomly distributed in mass ratios $q \in[1,50]$, spins $\chi_{1,2} \in [-0.99,0.99]$, initial eccentricity $e_0 \in [0.3,0.9]$ at a starting frequency of
$20$Hz for binaries with total mass $80 M_\odot$. In the generation of such dataset we do not find any waveform generation failure. 
However, lacking NR waveforms to compare against, we are not able to assess the accuracy and robustness of the model in this much larger region of the parameter space, so we recommend to use the model with caution for $e_0>0.3$.

Finally, we note that the region of parameter space with eccentricity up to $0.3$ at 20 Hz is 
is of significant astrophysical interest. In fact, it is expected that most of the GW events 
detected with ground-based detectors, such as LIGO, Virgo and KAGRA, have small eccentricities $\lesssim 0.1$
\cite{Wen:2002km,Samsing:2017xmd,Rodriguez:2017pec,Zevin:2021rtf}, which are typically defined at 10Hz.

The studies discussed in this section provide just a glance of all the internal checks 
performed to validate and implement the \texttt{SEOBNRv4EHM} waveform model in \texttt{LALSuite}
\cite{lalsuite}, so that it is available to the large GW community as a tool to carry out 
inference studies of GW signals.
 
\begin{table*}[!]
\caption{Summary of the eccentric NR simulations used in this work \cite{Hinder:2017sxy, Boyle:2019kee}. Each simulation is specified by the mass ratio $q= m_1/m_2\geq 1 $, z-component of the dimensionless spin vectors, $\chi_{1,2}$, 
the NR orbital frequency $\omega_{\text{orb,p}}$, the eccentricity measured from the NR orbital frequency $e_{\omega_{\text{orb,p}}}$, the $(2,2)$-mode frequency $\omega_{22,p}$, and the eccentricity measured from the $(2,2)$-mode frequency $e_{\omega_{22,p}}$, all evaluated at first periastron passage. For each simulation we report also the optimal values of the starting orbital frequency and eccentricity at periastron, ($\omega_{p}$, $e_{\omega_p}$),   for the  \texttt{SEOBNRv4E} and \texttt{TEOBResumSE} waveform models, as well as the maximum mismatch over the total mass range using such optimal values against the NR waveforms.}
\def\arraystretch{1.4}
\input{Tables/ecc_NR_table.tex}
\label{tab:tabNR}
\end{table*}

\subsection{Unfaithfulness between eccentric and quasi-circular waveforms}
\label{subsec:v4EHMvsv4HM}

The impact of eccentricity in GW data analysis --- for example Bayesian inference or GW searches --- has been 
investigated in the literature~\cite{Martel:1999tm,Huerta:2013qb,LIGOScientific:2019dag,Romero-Shaw:2019itr,Nitz:2019spj,Romero-Shaw:2020thy,Gayathri:2020coq,Ramos-Buades:2020eju,Favata:2021vhw,OShea:2021ugg, Romero-Shaw:2021ual,  Nitz:2021mzz, Lenon:2021zac}, but it is mostly restricted to inspiral-only eccentric waveforms. Here, we start to extend these studies to IMR eccentric waveforms, exploring the region of parameter space in which we expect biases in estimating the source's properties if quasi-circular--orbit waveforms were employed.

Using the \texttt{SEOBNRv4EHM} model, we compute the SNR-weighted
  unfaithfulness, averaged over the effective polarization angle and
  azimuthal angle of the signal, for an inclination angle of the source
  $\iota_s = \pi/3$ (Eq. \eqref{eq:eq18}), against the
  quasi-circular \texttt{SEOBNRv4HM} model in the following parameter
  space: $q \in [1,20]$, $\chi_{1,2}\in [-0.95,0.95]$, and $e_0 \in
  [0,0.3]$. As an example, we consider a total mass of $70 M_\odot$ and starting
  frequency of 20Hz. We fix the initial conditions at periastron
  $(\zeta_0 = 0)$ to reduce the dimensionality of the parameter space,
  and we set the starting frequency of the \texttt{SEOBNRv4HM}
  waveform, so that the length of the quasi-circular waveform is the
  same as the eccentric one produced with \texttt{SEOBNRv4EHM}. 

\begin{figure}[!]
\begin{center}
\includegraphics[width=1\columnwidth]{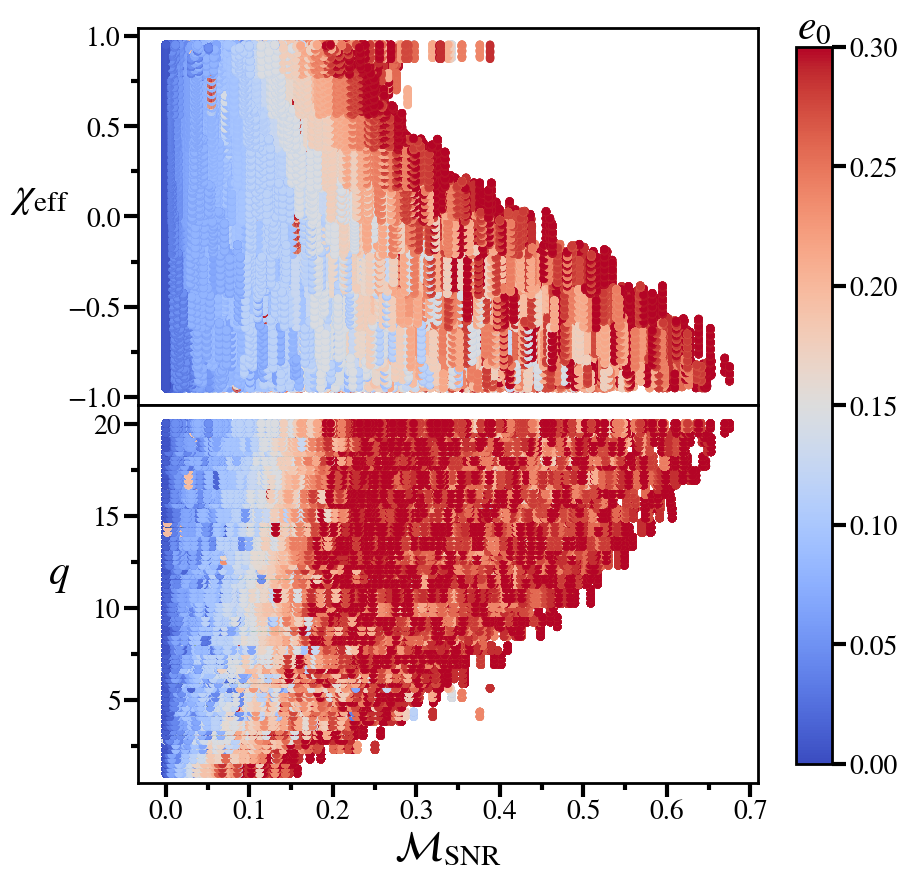}
\caption{SNR-weighted unfaithfulness, $\mathcal{M}_{\text{SNR}}=1-\mathcal{F}_{\text{SNR}}$, averaged over the effective polarization angle and azimuthal angle of the signal for an inclination angle of the signal $\iota_s = \pi/3$ as described in Eq. \eqref{eq:eq18}, between \texttt{SEOBNRv4EHM} and \texttt{SEOBNRv4HM} in the parameter space: $q \in [1,20]$, $\chi_{1,2}\in [-0.95,0.95]$, and $e_0 \in [0,0.3]$.
The total mass considered here is $70 M_\odot$, and the starting frequency is 20Hz. The
calculations are done with the Advanced LIGO's zero-detuned high-power design
sensitivity curve \cite{Barsotti:2018}. In the upper panel, we show the effective spin parameter, $\chi_{\text{eff}}$, defined in Eq. \eqref{eq:eq11.3}, as a function of SNR-weighted unfaithfulness. In the lower panel, we show the same quantity as in the upper one, but using mass ratio, $q$, in the $y$-axis. The color bar indicates the value of the initial eccentricity, $e_0$.
}
\label{fig:unfaithfulnessv4EHMv4HM}
\end{center}
\end{figure}

The unfaithfulness results are shown in Fig.~\ref{fig:unfaithfulnessv4EHMv4HM}. As expected, we observe that the unfaithfulness becomes increasingly large with eccentricity. We also appreciate a dependence of the results on the mass ratio and the effective-spin parameter. In the latter case, we observe that the unfaithfulness can be as large as  $\sim 70\%$ for large negative spins, while for positive spins the largest values occur at $\chi_{\text{eff}} \sim 0.95$. The unfaithfulness also shows a dependence on the mass ratio, with values up to $\sim 70\%$ for $q\sim 20$, while for comparable masses the unfaithfulness can be as large as $30\%$.

We note that large values of the unfaithfulness can imply large biases in source's parameters, if the quasi-circular models were employed in inference studies against eccentric GW signals. Moreover, large unfaithfulness can also lead to a loss in SNR, which can make the GW modeled searches suboptimal~\cite{Martel:1999tm,Huerta:2013qb}. The weighting of the unfaithfulness by the SNR (see Eq. \eqref{eq:eq18}), provides a conservative estimate of the upper limit of the fraction of detection volume lost. However, the unfaithfulness results presented here cannot be translated into an estimate of the sensitivity of a matched-filter search pipeline. This is because in a matched-filter search, the signal is compared against templates with different intrinsic parameters \cite{Allen:2005fk,Privitera:2013xza, Usman:2015kfa,messick2017analysis}, which is not the case of our unfaithfulness study, where we have fixed the intrinsic parameters of the signal and the template to be the same. More comprehensive studies with GW signals that cover a large portion of the parameter space should be pursued in the future to  assess the sensitivity of modeled searches to eccentric signals from BBHs with non-precessing spins, and quantify the biases in the estimation of the parameters if quasi-circular--orbit models were employed.

\section{Conclusions}
\label{sec:conclusions}

Working within the EOB framework, we have developed the multipolar eccentric
waveform model \texttt{SEOBNRv4EHM} for BBHs with non-precessing spins and 
multipoles $(l,|m|)=\{(2,2),(2,1),(3,3),(4,4),(5,5)\}$. The eccentric
waveform model is built upon the multipolar quasi-circular
\texttt{SEOBNRv4HM} model \cite{Bohe:2016gbl,Cotesta:2018fcv}. 
The inspiral waveform model \texttt{SEOBNRv4EHM} 
includes recently computed eccentric corrections up to the 2PN order
\cite{Khalil:2021txt}, including the spin-orbit and spin-spin interactions, 
in the factorized GW modes. By contrast the merger and ringdown description of 
the \texttt{SEOBNRv4EHM} model is not modified with respect to the one in the 
quasi-circular orbit case. Thus, we assume that the binary
circularizes by the time it merges, and this is in agreement with NR simulations 
for mild eccentricities~\cite{Hinder:2007qu,Hinder:2017sxy}.

We have generalized the eccentric initial conditions introduced in
Ref.~\cite{Khalil:2021txt} to include two eccentric parameters, the initial
eccentricity $e_0$, and the initial relativistic anomaly $\zeta_0$. Both parameters are 
specified at a certain starting frequency $\omega_0$ along the elliptical orbit. 
We note that when the binary starts its evolution at periastron or apastron, one
only needs to specify $(e_0,\omega_0)$, and this is the choice made 
in other eccentric EOB waveform models in the literature~
\cite{Liu:2021pkr, Chiaramello:2020ehz,Nagar:2021gss,Nagar:2021xnh,Placidi:2021rkh}.
The relativistic anomaly is degenerate with variations of the initial orbital frequency $\omega_0$ at fixed $e_0$. This fact is used to reduce the dimensionality of the parameter space when comparing EOB and NR waveforms  in Sec. \ref{sec:NReccentric}.
For applications like Bayesian-inference studies, having generic initial conditions becomes essential as the parameters of a binary system are inferred at a fixed reference frequency \cite{LIGOScientific:2020ibl}. Thus, for parameter-estimation studies, the starting frequency would be fixed, and the degeneracy between $\omega_0$ and  $\zeta_0$ can no longer be used to accurately sample the eccentric parameter space. 

We have also implemented the initial conditions for hyperbolic encounters
and dynamical-capture systems, expressing them in terms of the initial
angular momentum and energy at infinity~\cite{Damour:2014afa}. As an
example we have shown that the \texttt{SEOBNRv4EHM} model can
qualitatively reproduce the behavior of dynamical captures. We leave
to the future a quantitative and detailed study of the accuracy of the
model for hyperbolic encounters, including comparisons with NR
simulations of unbound systems.

Regarding the accuracy of our model, in the quasi-circular limit we have found that the unfaithfulness between the
(2,2)-mode only model, \texttt{SEOBNRv4E}, and the publicly available NR
waveforms used to construct and validate the underlying quasi-circular waveform
model, \texttt{SEOBNRv4}~\cite{Bohe:2016gbl}, is always smaller than $1\%$. 
When we include the higher order multipoles, $(l,|m|)=\{(2,1),(3,3),(4,4),(5,5)\}$, the unfaithfulness
averaged over sky positions, orientations and inclinations between
\texttt{SEOBNRv4EHM} and the same NR dataset used to validate
\texttt{SEOBNRv4HM}, is overall below $1\%$, with few configurations above
that threshold, but below $1.5 \%$, as in the case of \texttt{SEOBNRv4HM}
\cite{Cotesta:2018fcv}. Thus, the multipolar eccentric 
model has an accuracy comparable to the underlying quasi-circular model in the
zero eccentricity limit.

To asses the accuracy of the model in the eccentric case, we have
developed a maximization procedure of the unfaithfulness to estimate the optimal values of
eccentricity and starting frequency. For the $(2,2)$-mode waveforms, we have 
found that the unfaithfulness of the \texttt{SEOBNRv4E} model 
against the eccentric NR waveforms at our disposal, which have eccentricity $\lesssim 0.3$, 
is always smaller than $1\%$. We have also used another eccentric EOB waveform model
\texttt{TEOBResumSE} to compute the unfaithfulness against eccentric
NR waveforms, and we have found that the unfaithfulness is also always
$<1\%$. Overall we find that the unfaithfulness of \texttt{SEOBNRv4E} model 
is smaller than the public version of the \texttt{TEOBResumSE} model, at the time
of this publication, for the NR dataset at our disposal. We note that
in order to set more stringent constraints on the accuracy of both
models, comparisons against larger datasets of eccentric NR simulations
are required.

Considering  that the accuracy of the \texttt{SEOBNRv4EHM} model against 
NR simulations can currently be investigated only up to eccentricity $0.3$, we 
have assessed the smoothness and robustness of the \texttt{SEOBNRv4EHM} 
model in the parameter space $\chi_{1,2}\in[0.,0.99]$, $q\in[1,50]$ and $e= \in[0,0.3]$.  
We have found that some configurations when the spins are large and positive, 
notably in the range $\chi_{1,2}\in[0.95,0.99]$, lead to spurious oscillations 
in the amplitude and frequency close to merger. This is due to the suboptimal 
procedure used by the \texttt{SEOBNRv4EHM} model to transit from the late inspiral 
to the merger and ringdown. Furthermore, the \texttt{SEOBNRv4EHM} can be generated 
also for eccentricity larger than $0.3$, however we caution its use for large 
eccentricities, especially beyond the inspiral phase, since the model has been 
built under the assumption that the binary circularizes. We emphasize 
that current GW detectors, such as LIGO, Virgo and KAGRA 
will be able to detect eccentric GW events with mild eccentricities~
\cite{Samsing:2017xmd,Rodriguez:2017pec}. Thus, having a waveform model 
that can grasp the main characteristic of eccentric signals up to eccentricity 
$0.3$ is valuable --- for example the \texttt{SEOBNRv4EHM} model could 
be employed to search for eccentric signals in the LIGO and Virgo data 
and to infer the properties of the eccentric sources.

We remark that in this first eccentric EOBNR model, we have only
included the eccentric corrections up to 2PN order derived in Ref.~\cite{Khalil:2021txt} 
to the factorized GW modes, while we have kept the conservative and
dissipative dynamics the same as in the quasi-circular \texttt{SEOBNRv4HM} model. 
Preliminary comparisons with a larger set of NR simulations are indicating that better accuracy can be achieved when including the eccentric corrections~\cite{Khalil:2021txt} 
also in the RR forces. These improvements will be included in the next generation of the 
EOBNR waveform models currently under construction, that is the \texttt{SEOBNRv5} model.

We leave to the near future the development of a reduced-order-model (ROM)
\cite{Purrer:2014fza,Yun:2021jnh} version of the \texttt{SEOBNRv4EHM} model, so 
that it can efficiently be used for inference studies on current GW catalogs and 
future observations with the LIGO, Virgo and KAGRA detectors. We also plan to extend 
the \texttt{SEOBNRv4EHM} model to precessing binaries, including the eccentric corrections 
to the waveform modes of the quasi-circular spin-precessing \text{SEOBNRv4PHM} 
model~\cite{Pan:2013rra,Babak:2016tgq,Ossokine:2020kjp}.

\section*{Acknowledgments}
\label{acknowledgements}

It is a pleasure to thank Jan Steinhoff and Justin Vines for helpful
discussions. The computational work for this manuscript was carried out on the
computer cluster \texttt{Minerva} at the Max Planck Institute for Gravitational
Physics in Potsdam, and on the cluster \texttt{CIT} provided by the LIGO Laboratory
and supported by the National Science Foundation Grants PHY0757058 and PHY-0823459.
LIGO is funded by the U.S. National Science Foundation.


\appendix


\section{Eccentric corrections to the waveform modes} 
\label{app:EccModes}

In this appendix, we list the eccentric corrections to the waveform modes obtained in Ref.~\cite{Khalil:2021txt}. These corrections are written in terms of the dynamical quantities $r,p_r$ and $\dot p_r$. To ease the notation, we define
\begin{equation}
v_{\dot{\phi}} \equiv \frac{(\dot{p}_r r^2+1)^{1/6}}{\sqrt{r}},
\end{equation} 
which is $(\dot{\phi})^{1/3}$ at leading PN order for generic orbits, and it  
reduces to $v_\omega \equiv \omega^{1/3}$ in the circular-orbit limit. We also define the anti-symmetric mass ratio $\delta\equiv (m_1-m_2)/M$ and the spin combinations
\begin{equation}
\chi_S = \frac{1}{2}(\chi_1 + \chi_2), \qquad
\chi_A = \frac{1}{2}(\chi_1 - \chi_2).
\end{equation}

We expand the eccentric part of the leading-order tail term, $T_{lm}^\text{ecc}$, in Eq.~\eqref{eq:eq5}, in powers of the eccentricity 
up to $\mathcal{O}(e^6)$, and express it in terms of the dynamical quantities $p_r$ and $\dot p_r$ as described in Ref.~\cite{Khalil:2021txt} 
using the Keplerian parametrization. For the $\{(2,2),(2,1),(3,3)\}$ modes, we obtain through 2PN order 
\begin{widetext}
\begin{align}
T_{22}^\text{ecc} &= -\frac{\pi }{4c^3 r} \bigg[
\left(4 r^{3/2} \dot{p}_r+6 i p_r\right)
+\left(2 \sqrt{r} p_r^2+i r^2 p_r \dot{p}_r\right)
+ \left(\frac{5}{4} r^{5/2} p_r^2 \dot{p}_r+\frac{1}{12} r^{11/2} \dot{p}_r^3-\frac{3}{4} i r^4 p_r \dot{p}_r^2+\frac{5}{12} i r p_r^3\right) \nonumber\\
&\qquad
+ \left(\frac{5}{4} r^{3/2} p_r^4-r^{9/2} p_r^2 \dot{p}_r^2-\frac{1}{4} r^{15/2} \dot{p}_r^4+\frac{1}{8} i r^6 p_r \dot{p}_r^3-\frac{15}{8} i r^3 p_r^3 \dot{p}_r\right) \nonumber\\
&\qquad
+\left(-\frac{115}{48}  r^{7/2} p_r^4 \dot{p}_r+\frac{101}{96} r^{13/2} p_r^2 \dot{p}_r^3+\frac{29}{120} r^{19/2} \dot{p}_r^5-\frac{11}{96} i r^8 p_r \dot{p}_r^4+\frac{53}{24} i r^5 p_r^3 \dot{p}_r^2-\frac{589}{480} i r^2 p_r^5\right) \nonumber\\
&\qquad
+\left(-\frac{329}{480} r^{5/2} p_r^6+\frac{83}{32} r^{11/2} p_r^4 \dot{p}_r^2-\frac{257}{192} r^{17/2} p_r^2 \dot{p}_r^4-\frac{7}{36} r^{23/2} \dot{p}_r^6+\frac{223}{960} i r^{10} p_r \dot{p}_r^5-\frac{181}{72} i r^7 p_r^3 \dot{p}_r^3+\frac{111}{64} i r^4 p_r^5 \dot{p}_r\right) \nonumber\\
&\qquad
+ \mathcal{O}(p_r,\dot{p}_r)^8
\bigg], \\
T_{21}^\text{ecc} &= -\frac{\pi }{4c^4 r} \bigg[
8 i p_r
+\left(-r^{7/2}\dot{p}_r^2-2 i r^2 p_r \dot{p}_r+\sqrt{r} p_r^2\right)
+ \left(\frac{3}{2} r^{5/2} p_r^2 \dot{p}_r+\frac{5}{6} r^{11/2} \dot{p}_r^3+\frac{2}{3} i r p_r^3\right) \nonumber\\
&\qquad
+ \left(\frac{7}{16} r^{3/2} p_r^4-\frac{9}{4} r^{9/2} p_r^2 \dot{p}_r^2-\frac{7}{16} r^{15/2} \dot{p}_r^4+\frac{5}{4} i r^6 p_r \dot{p}_r^3-\frac{5}{4} i r^3 p_r^3 \dot{p}_r\right) \nonumber\\
&\qquad
+ \left(-\frac{79}{96}  r^{7/2} p_r^4 \dot{p}_r+2 r^{13/2} p_r^2 \dot{p}_r^3+\frac{33}{160} r^{19/2} \dot{p}_r^5-\frac{11}{8} i r^8 p_r \dot{p}_r^4+\frac{19}{12} i r^5 p_r^3 \dot{p}_r^2-\frac{19}{120} i r^2 p_r^5\right)\nonumber\\
&\qquad
+ \left(-\frac{59}{320} r^{5/2} p_r^6+\frac{145}{384} r^{11/2} p_r^4 \dot{p}_r^2-\frac{487}{384} r^{17/2} p_r^2 \dot{p}_r^4-\frac{161 r^{23/2} \dot{p}_r^6}{1440}+\frac{511}{480} i r^{10} p_r \dot{p}_r^5-\frac{115}{144} i r^7 p_r^3 \dot{p}_r^3+\frac{61}{160} i r^4 p_r^5 \dot{p}_r\right)\nonumber\\
&\qquad
+ \mathcal{O}(p_r,\dot{p}_r)^8
\bigg], \\
T_{33}^\text{ecc} &= -\frac{\pi }{81c^4 r} \bigg[
\left(90 r^{3/2} \dot{p}_r+180 i p_r\right)
+ \left(-11 r^{7/2} \dot{p}_r^2+22 i r^2 p_r \dot{p}_r+56 \sqrt{r} p_r^2\right)
+ \left(2 r^{5/2} p_r^2 \dot{p}_r+\frac{331}{36} r^{11/2} \dot{p}_r^3-\frac{8}{3} i r^4 p_r \dot{p}_r^2+\frac{34}{9} i r p_r^3\right) \nonumber\\
&\qquad
+\left(\frac{26161 r^{3/2} p_r^4}{5184}+\frac{10895}{864} r^{9/2} p_r^2 \dot{p}_r^2-\frac{33191 r^{15/2} \dot{p}_r^4}{5184}-\frac{2383 i r^6 p_r \dot{p}_r^3}{1296}+\frac{14671 i r^3 p_r^3 \dot{p}_r}{1296}\right) \nonumber\\
&\qquad
+\left(\frac{3165361 r^{7/2} p_r^4 \dot{p}_r}{93312}-\frac{1286269 r^{13/2} p_r^2 \dot{p}_r^3}{46656}+\frac{2170663 r^{19/2} \dot{p}_r^5}{466560}+\frac{17347 i r^8 p_r \dot{p}_r^4}{2916}-\frac{252673 i r^5 p_r^3 \dot{p}_r^2}{5832}+\frac{590461 i r^2 p_r^5}{58320}\right) \nonumber\\
&\qquad
+\bigg(\frac{269913797 r^{5/2} p_r^6}{16796160}-\frac{1192721 r^{11/2} p_r^4 \dot{p}_r^2}{15552}+\frac{49153087 r^{17/2} p_r^2 \dot{p}_r^4}{1119744}-\frac{6543119 r^{23/2} \dot{p}_r^6}{1679616}-\frac{10372969 i r^{10} p_r \dot{p}_r^5}{933120} \nonumber\\
&\quad\qquad
+\frac{63206059 i r^7 p_r^3 \dot{p}_r^3}{839808}-\frac{27211231 i r^4 p_r^5 \dot{p}_r}{559872}\bigg)
+ \mathcal{O}(p_r,\dot{p}_r)^8
\bigg].
\end{align}

For the eccentric term $f_{lm}^\text{ecc}$ in Eq.~\eqref{eq:eq5}, and the modes $\{(2,2),(2,1),(3,3),(4,4),(5,5)\}$, we obtain

\begin{align}
f_{22}^\text{ecc} &= \frac{1-r p_r^2+r^3 v_{\dot{\phi}}^6-2 r v_{\dot{\phi}}^2}{2 r v_{\dot{\phi}}^2}+i r v_{\dot{\phi}} p_r 
+ \frac{1}{84c^2r^5v_{\dot{\phi}}^8} \bigg\lbrace
-14 (\nu +1)+ i r^2 v_{\dot{\phi}}^3 p_r \left[14 (\nu +1)+r^3 v_{\dot{\phi}}^6 \left(-101 \nu +(41 \nu -37) r^3 v_{\dot{\phi}}^6-209\right)\right]\nonumber\\
&\qquad 
+r^2 p_r^4 \left[(29-10 \nu ) r^3 v_{\dot{\phi}}^6-7 (\nu -3)\right]
+r^3 v_{\dot{\phi}}^6 \left[63 \nu +r^2 v_{\dot{\phi}}^4 \left(-110 \nu +(31 \nu -8) r^4 v_{\dot{\phi}}^8+(30 \nu -59) r v_{\dot{\phi}}^2+172\right)-91\right]\nonumber\\
&\qquad
+p_r^2 r \left[21 \nu +3 r^3 v_{\dot{\phi}}^6 \left(3 \nu +7 (\nu +1) r^3 v_{\dot{\phi}}^6+62\right)-7\right] 
+i r^3 v_{\dot{\phi}}^3 p_r^3 \left[(41 \nu -37) r^3 v_{\dot{\phi}}^6-7 (\nu -3)\right] \bigg\rbrace \nonumber\\
&\quad
+\frac{1}{6048c^4 r^9 v_{\dot{\phi}}^{14}} \bigg\lbrace
3 i (103 \nu^2 -700\nu+127) r^{14} v_{\dot{\phi}}^{27} p_r+6 r^{12} v_{\dot{\phi}}^{24} \left[318-4 \nu  (96 \nu +265)+( 40 \nu^2 +20 \nu-71) r p_r^2\right] \nonumber\\
&\qquad
+6 i r^{11} v_{\dot{\phi}}^{21} p_r \left[80+1531\nu-640 \nu^2 + (349 -143 \nu^2 -724\nu) r p_r^2\right]
+24 (55 \nu -86) r^8 v_{\dot{\phi}}^{16} \left[1-\nu +(\nu +1) r p_r^2\right] \nonumber\\
&\qquad
+6 r^9 v_{\dot{\phi}}^{18} \left[-616+955\nu-409 \nu^2+(17 \nu^2 +193\nu-340) r^2 p_r^4-(1181 \nu^2 +1277\nu+279) r p_r^2\right] \nonumber\\
&\qquad
-105 i r^2 v_{\dot{\phi}}^3 p_r \left[(\nu -3) r p_r^2-2 (\nu +1)\right]^2
+24 (55 \nu -86) r^5 v_{\dot{\phi}}^{10} \left[(\nu -3) r p_r^2-2 (\nu +1)\right] \nonumber\\
&\qquad
+i r^8 v_{\dot{\phi}}^{15} p_r \left[13 \nu  (3598-517 \nu )-3 (389 \nu^2 +748\nu-571) r^2 p_r^4+12 (281 \nu^2 -349\nu+803) r p_r^2-9221\right] \nonumber\\
&\qquad
+6 i r^5 v_{\dot{\phi}}^9 p_r \left[50 \nu ^2-872 \nu +(22 \nu^2 +55\nu-216) r^2 p_r^4-(69 \nu^2 +674\nu+953) r p_r^2-166\right] \nonumber\\
&\qquad
+6 r^3 v_{\dot{\phi}}^6 \left[28 (17\nu-16 \nu^2+6)+(43 \nu^2 +13\nu-279) r^3 p_r^6-(243 \nu^2 +325\nu+1253) r^2 p_r^4+2 (269 \nu^2 -446\nu+237) r p_r^2\right] \nonumber\\
&\qquad
+6 r^6 v_{\dot{\phi}}^{12} \left[2 \nu  (733 \nu -620)+(80 \nu^2 +130\nu-277) r^3 p_r^6+\left(-294 \nu ^2+380 \nu -1963\right) r^2 p_r^4-2 (  205 \nu^2 +2604\nu+93) r p_r^2+26\right]\nonumber\\
&\qquad
-168 \left(r p_r^2-1\right) \left[(\nu -3) r p_r^2-2 (\nu +1)\right]^2+6 (103 \nu^2 -43\nu-8) r^{15} v_{\dot{\phi}}^{30}+24 (\nu +1) (55 \nu -86) r^{11} v_{\dot{\phi}}^{22}
\bigg\rbrace \nonumber\\
&\quad
+ \frac{1}{6 c^3 r^4 v_{\dot{\phi}}^5} \bigg\lbrace
\chi _S \left[2-\nu +i (5 \nu -8) r^2 v_{\dot{\phi}}^3 p_r+(\nu -2) r p_r^2+r^3 v_{\dot{\phi}}^6 \left(9 \nu -8 (\nu -1) r v_{\dot{\phi}}^2-10\right)\right] \nonumber\\
&\qquad
-2 \delta  \chi _A \left[-1+4 i r^2 v_{\dot{\phi}}^3 p_r+r p_r^2+r^3 v_{\dot{\phi}}^6 \left(5-4 r v_{\dot{\phi}}^2\right)\right]
\bigg\rbrace \nonumber\\
&\quad
+ \frac{1}{6 c^4 r^6 v_{\dot{\phi}}^8} \bigg\lbrace
2 \delta  \chi _A \chi _S \bigg[i (2 \nu -1) r^3 v_{\dot{\phi}}^3 p_r^3+r p_r^2 \left(\nu +5 (2 \nu -1) r^3 v_{\dot{\phi}}^6+1\right)+(2 \nu -1) r^2 p_r^4-3 i p_r \left(2 (2 \nu -1) r^5 v_{\dot{\phi}}^9-\nu  r^2 v_{\dot{\phi}}^3\right) \nonumber\\
&\quad\qquad
-3 \left(r^3 v_{\dot{\phi}}^6-1\right) \left((2 \nu +1) r^3 v_{\dot{\phi}}^6-\nu \right)\bigg] 
+\chi _S^2 \bigg[r p_r^2 \left(-2 \nu ^2+2 \nu -5 (1-2 \nu )^2 r^3 v_{\dot{\phi}}^6+1\right)-i (1-2 \nu )^2 r^3 v_{\dot{\phi}}^3 p_r^3-(1-2 \nu )^2 r^2 p_r^4 \nonumber\\
&\quad\qquad
+6 i p_r \left((1-2 \nu )^2 r^5 v_{\dot{\phi}}^9-(\nu -1) \nu  r^2 v_{\dot{\phi}}^3\right)+3 \left(r^3 v_{\dot{\phi}}^6-1\right) \left((4 \nu ^2-4 \nu -1) r^3 v_{\dot{\phi}}^6-2 (\nu -1) \nu \right)\bigg] \nonumber\\
&\qquad
+ \chi _A^2 (4 \nu -1) r\left[-6 i r^4 v_{\dot{\phi}}^9 p_r+p_r^2 \left(5 r^3 v_{\dot{\phi}}^6-1\right)+i r^2 v_{\dot{\phi}}^3 p_r^3+r p_r^4+3 r^2 v_{\dot{\phi}}^6 \left(r^3 v_{\dot{\phi}}^6-1\right)\right]
\bigg\rbrace + \mathcal{O}\left(\frac{1}{c^5}\right), \\
f_{21}^\text{ecc} &= \frac{1-r^2 v_{\dot{\phi}}^4}{r^2 v_{\dot{\phi}}^4} - \frac{1}{42c^2 r^6 v_{\dot{\phi}}^{10}} \left\lbrace
28 (\nu +1)+3 i (12 \nu -83) r^5 v_{\dot{\phi}}^9 p_r+r^3 v_{\dot{\phi}}^6 \left(82-52 \nu +(19 \nu +106) r p_r^2\right)-14 (\nu -3) r p_r^2+2 (12 \nu -55) r^6 v_{\dot{\phi}}^{12}
\right\rbrace \nonumber\\
&\quad
+ \frac{1}{4c \delta  r^4 v_{\dot{\phi}}^7} \left[\chi _A \left(6 r^4 v_{\dot{\phi}}^8-6\right)+6 \delta  \left(r^4 v_{\dot{\phi}}^8-1\right) \chi _S\right]
+ \frac{1}{336 c^3\delta  r^8 v_{\dot{\phi}}^{13}} \bigg\lbrace
\chi _A \bigg[588 (\nu +1)+24 i (104 \nu -147) r^5 v_{\dot{\phi}}^9 p_r+42 (\nu -1) r^7 v_{\dot{\phi}}^{14}\nonumber\\
&\quad\qquad
-6 p_r^2 \left(7 (\nu +1) r^8 v_{\dot{\phi}}^{14}+7 (\nu -3) r^5 v_{\dot{\phi}}^8-(141 \nu +203) r^4 v_{\dot{\phi}}^6+49 (\nu -3) r\right)
-42 (\nu +1) r^{10} v_{\dot{\phi}}^{20}-4 (131 \nu +427) r^8 v_{\dot{\phi}}^{16} \nonumber\\
&\quad\qquad
+30 (121 \nu -35) r^6 v_{\dot{\phi}}^{12}
+84 (\nu +1) r^4 v_{\dot{\phi}}^8+2 (1085-1889 \nu ) r^3 v_{\dot{\phi}}^6\bigg]
-2 \delta  \chi _S \bigg[-294 (\nu +1)+36 i (2 \nu +49) r^5 v_{\dot{\phi}}^9 p_r \nonumber\\
&\quad\qquad
+3 p_r^2 \left(7 (\nu +1) r^8 v_{\dot{\phi}}^{14}+7 (\nu -3) r^5 v_{\dot{\phi}}^8-(69 \nu +203) r^4 v_{\dot{\phi}}^6+49 (\nu -3) r\right)
+21 (\nu +1) r^{10} v_{\dot{\phi}}^{20}+2 (79 \nu +427) r^8 v_{\dot{\phi}}^{16} \nonumber\\
&\quad\qquad
-21 (\nu -1) r^7 v_{\dot{\phi}}^{14}
+15 (35-33 \nu ) r^6 v_{\dot{\phi}}^{12}-42 (\nu +1) r^4 v_{\dot{\phi}}^8+(673 \nu -1085) r^3 v_{\dot{\phi}}^6\bigg]
\bigg\rbrace + \mathcal{O}\left(\frac{1}{c^4}\right),
\\
f_{33}^\text{ecc} &= \frac{2 }{9 r v_{\dot{\phi}}^3}\left[3 i r^3 v_{\dot{\phi}}^6 p_r-3 r^2 v_{\dot{\phi}}^3 p_r^2-i r p_r^3+2 i p_r+r^4 v_{\dot{\phi}}^9-r v_{\dot{\phi}}^3\right]
+ \frac{1}{162c^2 r^5 v_{\dot{\phi}}^9} \bigg\lbrace
-36 (\nu -2) r^6 v_{\dot{\phi}}^9 p_r^4
+6 r^5 v_{\dot{\phi}}^9 p_r^2 \left[18 \nu -(\nu -11) r^3 v_{\dot{\phi}}^6+57\right]
 \nonumber\\
&\qquad
+6 i r p_r^3 \left[6 (\nu -1)+r^3 v_{\dot{\phi}}^6 \left(-4 \nu +2 (5 \nu -1) r^3 v_{\dot{\phi}}^6+35\right)\right]
+3 r^4 v_{\dot{\phi}}^9 \left[56 \nu +r^3 v_{\dot{\phi}}^6 \left(42 \nu +2 (5 \nu -1) r^3 v_{\dot{\phi}}^6-39\right)+27 (7-4 \nu ) r v_{\dot{\phi}}^2-148\right] \nonumber\\
&\qquad
+i p_r \left[r^3 v_{\dot{\phi}}^6 \left(100 \nu +9 r^3 v_{\dot{\phi}}^6 \left(-23 \nu +(7 \nu -5) r^3 v_{\dot{\phi}}^6-29\right)-218\right)-36 (\nu +1)\right]
-3 i r^2 p_r^5 \left[3 (\nu -3)+(\nu -11) r^3 v_{\dot{\phi}}^6\right]
\bigg\rbrace \nonumber\\
&\quad
\frac{1}{18c^3 \delta  r^4 v_{\dot{\phi}}^6} \bigg\lbrace
\chi _A \left[2 i p_r \left(-16 \nu +(101 \nu -24) r^3 v_{\dot{\phi}}^6+4\right)+2 (6-25 \nu ) r^2 v_{\dot{\phi}}^3 p_r^2+4 i (4 \nu -1) r p_r^3-4 (5 \nu -1) r v_{\dot{\phi}}^3 \left(r^3 v_{\dot{\phi}}^6-1\right)\right] \nonumber\\
&\qquad
+2 \delta  \chi _S \left[i p_r \left(-2 \nu +(17 \nu -24) r^3 v_{\dot{\phi}}^6+4\right)+2 (3-2 \nu ) r^2 v_{\dot{\phi}}^3 p_r^2+i (\nu -2) r p_r^3-(3 \nu -2) r v_{\dot{\phi}}^3 \left(r^3 v_{\dot{\phi}}^6-1\right)\right]
\bigg\rbrace + \mathcal{O}\left(\frac{1}{c^4}\right), \\
f_{44}^\text{ecc} &= \frac{1}{64 r^2 v_{\dot{\phi}}^4} \left\lbrace7+r \left[24 i r^4 v_{\dot{\phi}}^9 p_r+3 r^2 v_{\dot{\phi}}^6 \left(17-12 r p_r^2\right)-6 i r v_{\dot{\phi}}^3 p_r \left(4 r p_r^2-9\right)+6 p_r^2 \left(r p_r^2-3\right)+6 r^5 v_{\dot{\phi}}^{12}-64 r v_{\dot{\phi}}^4\right]\right\rbrace \nonumber\\
&\quad
+ \frac{1}{42240 (3 \nu -1)c^2 r^6 v_{\dot{\phi}}^{10}} \bigg\lbrace
-3 r^9 v_{\dot{\phi}}^{18} \left[60 \nu  (889-636 \nu )+20 (\nu  (321 \nu -926)+238) r p_r^2-10481\right] \nonumber\\
&\qquad
+120 i (267 \nu^2 -278\nu+49) r^{11} v_{\dot{\phi}}^{21} p_r
+6 i r^8 v_{\dot{\phi}}^{15} p_r \left[8033-5 \nu  (4551 \nu +3622)+1320 (\nu +1) (3 \nu -1) r p_r^2\right] \nonumber\\
&\qquad
+r^6 v_{\dot{\phi}}^{12} \left[-780 \nu  (234 \nu -289)-60 (519 \nu^2 -794\nu+172) r^2 p_r^4+6 (10 \nu  (2265 \nu +1882)-9847) r p_r^2-46063\right] \nonumber\\
&\qquad
-6 i r^5 v_{\dot{\phi}}^9 p_r \left[5 \nu  (12748-2727 \nu )+20 (69 \nu^2 -410\nu+115) r^2 p_r^4+(10277-5 \nu  (525 \nu +5294)) r p_r^2-20789\right] \nonumber\\
&\qquad
+20 r^3 v_{\dot{\phi}}^6 \left[3315 \nu^2 -2644\nu-3 (15 \nu^2 +238\nu-74) r^3 p_r^6+3 (339 \nu^2 -1274\nu+422) r^2 p_r^4-2 (1797 \nu^2 -4910\nu+1510) r p_r^2+553\right] \nonumber\\
&\qquad
-330 i (3 \nu -1) r^2 v_{\dot{\phi}}^3 p_r \left(4 r p_r^2-9\right) \left[(\nu -3) r p_r^2-2 (\nu +1)\right]
+220 (3 \nu -1) \left[6 r \left(r p_r^2-3\right) p_r^2+7\right] \left[(\nu -3) r p_r^2-2 (\nu +1)\right] \nonumber\\
&\qquad
+60 (183 \nu^2 -106\nu+8) r^{12} v_{\dot{\phi}}^{24}
\bigg\rbrace + \mathcal{O}\left(\frac{1}{c^3}\right), \\
f_{55}^\text{ecc} &= \frac{1}{625 r^2 v_{\dot{\phi}}^5} \bigg\lbrace
120 i r^6 v_{\dot{\phi}}^{12} p_r+3 r^4 v_{\dot{\phi}}^9 \left(143-80 r p_r^2\right)-48 i r^3 v_{\dot{\phi}}^6 p_r \left(5 r p_r^2-13\right)+4 r v_{\dot{\phi}}^3 \left(30 r^2 p_r^4-99 r p_r^2+43\right)
+24 r^7 v_{\dot{\phi}}^{15}-625 r^2 v_{\dot{\phi}}^5 \nonumber\\
&\qquad
+2 i p_r \left[12 r \left(r p_r^2-4\right) p_r^2+41\right]
\bigg\rbrace + \mathcal{O}\left(\frac{1}{c}\right).
\end{align}
\end{widetext}
For binaries of equal masses ($\delta=0,\,\nu=1/4$), the leading-PN order of the odd-$m$ modes is proportional to $\delta$, which cancels with the denominator of $\chi_A/\delta$ in the above expressions for the $(2,1)$ and $(3,3)$ modes, leading to
\begin{align}
f_{21}^{\text{ecc},\delta=0} &= \frac{3  \chi _A }{2c v_{\dot{\phi}}^7}\left(v_{\dot{\phi}}^8-\frac{1}{r^4}\right) \nonumber\\
&
+\frac{\chi _A }{224c^3 r^8 v_{\dot{\phi}}^{13}} \bigg[r p_r^2 \left(35 r^7 v_{\dot{\phi}}^{14}-77 r^4 v_{\dot{\phi}}^8-953 r^3 v_{\dot{\phi}}^6-539\right) \nonumber\\
&\quad
+1936 i r^5 v_{\dot{\phi}}^9 p_r+35 r^{10} v_{\dot{\phi}}^{20}+1226 r^8 v_{\dot{\phi}}^{16}+21 r^7 v_{\dot{\phi}}^{14} \nonumber\\
&\quad+95 r^6 v_{\dot{\phi}}^{12}-70 r^4 v_{\dot{\phi}}^8-817 r^3 v_{\dot{\phi}}^6-490\bigg], \\
f_{33}^{\text{ecc},\delta=0} &= \frac{\chi _A }{36 c^3 r^3 v_{\dot{\phi}}^3} \left[2-r \left(-5 i r v_{\dot{\phi}}^3 p_r+p_r^2+2 r^2 v_{\dot{\phi}}^6\right)\right].
\end{align}
Since the eccentric correction to the $(5,5)$-mode does not depend on spin at this order, it goes to zero for equal masses.

\section{Implementation of the orbit averaging procedure}
\label{sec:AppendixA}

In this appendix, we describe in detail the  orbit averaging  procedure that we have applied in Sec. \ref{sec:WaveformModes}  to the instantaneous NQC functions of the waveform.

According to Eq.~\eqref{eq:eq9}, to orbit average a dynamical quantity we need to define the times, $t_i$, which identify successive orbits in the evolution. In our implementation, we use the local 
maxima and employ a simple algorithm that compares each element of the time series with the two closest neighbors. However, these values may strongly depend on the
specified time step, thus, in order to reduce such dependence we further
compute the parabola passing through these three points $\{ (t_{i-1},X_{i-1}),
(t_i,X_{i}), (t_{i+1},X_{i+1})\}$, and obtain its maxima analytically, 
\begin{equation}
\begin{split}
f(t) &=a t^2 +b t +c , \quad t_{\text{max}} = -b/2a, \\
\Delta &= (t_i- t_{i+1})(t_i - t_{i-1})(t_{i+1} - t_{i-1})\\
a  & =\frac{1}{\Delta}\left[X_{i-1}(t_i - t_{i+1}) + X_i(t_{i+1} - t_{i-1}) +  X_{i+1}(-t_i + t_{i-1}))\right],  \\
b & = \frac{1}{\Delta}\left[X_{i-1}(-t_i^2 + t_{i+1}^2) + X_{i+1} (t_i^2  -t_{i-1}^2) + X_i (-t_{i+1}^2 + t_{i-1}^2)\right], \\
c & = \frac{1}{\Delta}\left[ (X_{i-1}(t_i - t_{i+1})t_i t_{i+1} + X_i ( t_{i+1} - t_{i-1})t_{i-1}t_{i+1}  \right. \\
    &\left.  + X_{i+1}t_i t_{i-1}(-t_i+ t_{i-1}))\right],
\end{split}
\label{eq:eq11.1}
\end{equation}
where $t_{\text{max}}$ is the solution of $df(t)/dt = 0$. The found maxima, and their corresponding times, $\{t_i, X_i \}$,  are then used in Eq. \eqref{eq:eq9} to compute the orbit-average quantity $\overline{X}_i$, and the intermediate times, $\overline{t}_i=(t_{i+1}+t_i)/2$ \cite{Lewis:2016lgx}, are associated to each   $\overline{X}_i$.

As our dynamical quantity, we use $p_{r_*}$, but when the eccentricity $e_0<0.1$, 
we switch to $\dot{p}_{r*}$, because we find that we can reliably extract the 
maxima in this quantity even for very small eccentricities ($e_0<0.01$). This is 
due to the fact that the time derivative enhances the effects of the eccentric
oscillations.

\begin{figure}[htbp!]
\begin{center}
\includegraphics[width=1.\columnwidth]{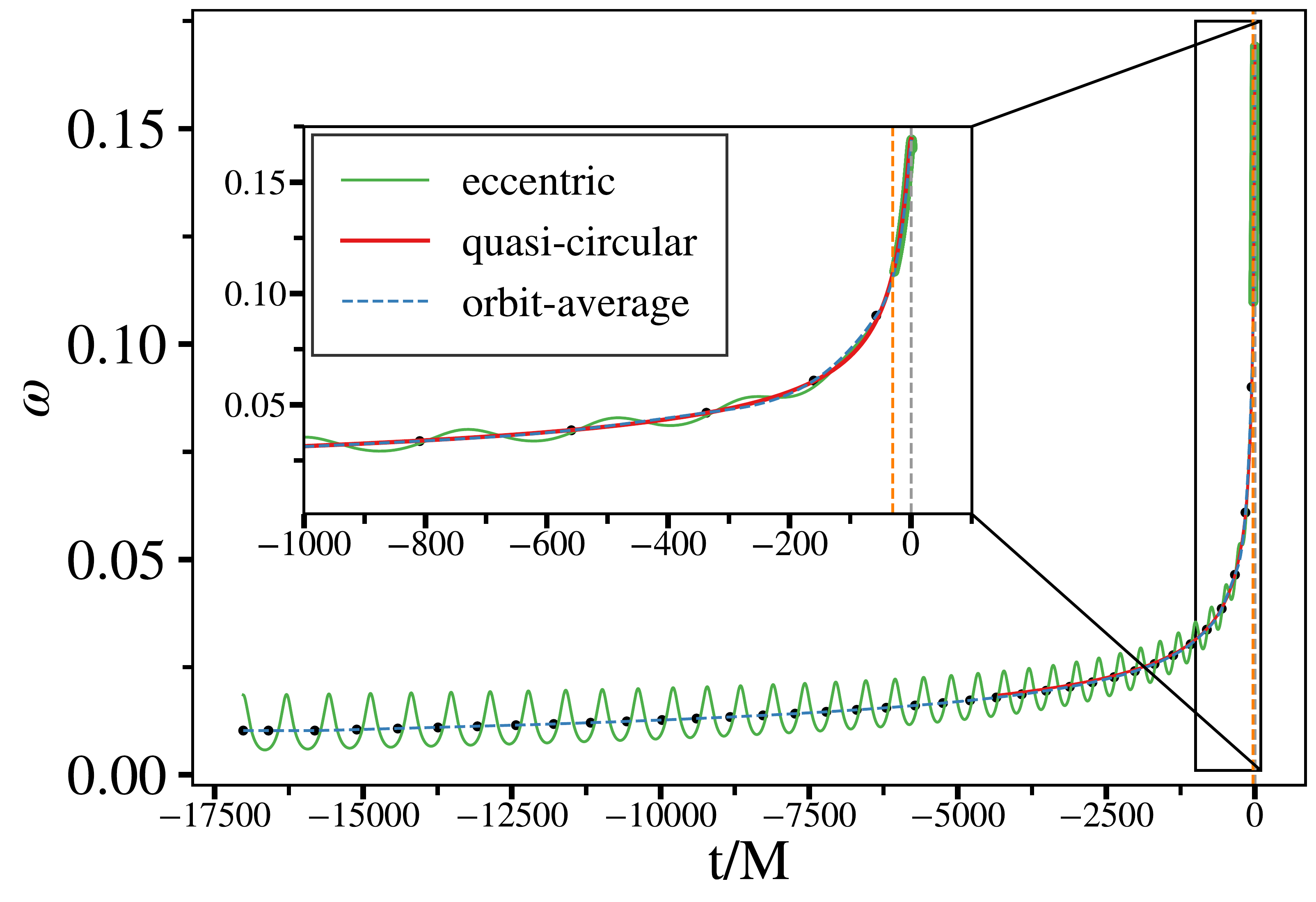}
\includegraphics[width=1.\columnwidth]{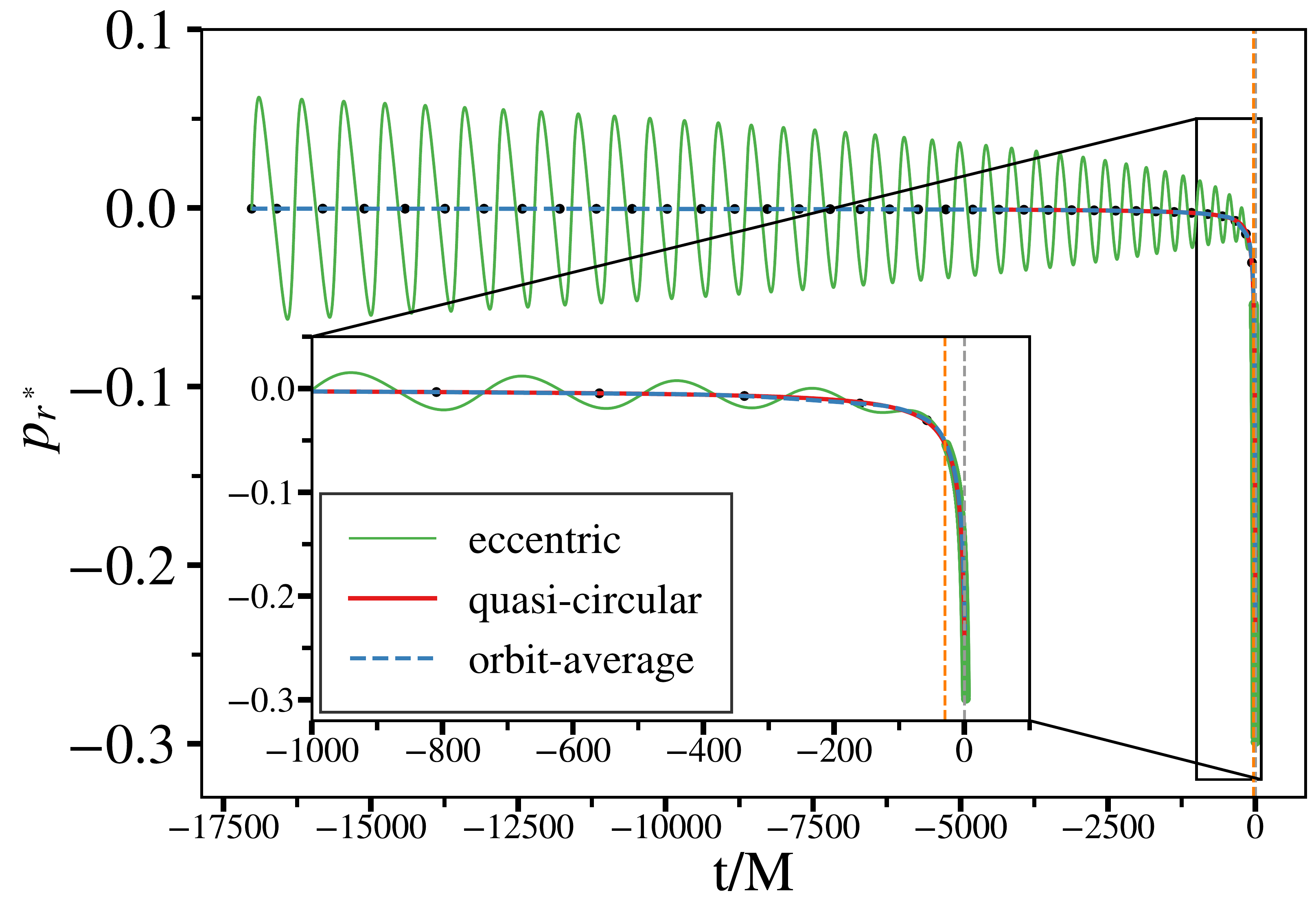}
\caption{Time evolution of the orbital frequency $\omega$ (upper panel), and the radial momentum $p_{r*}$ (lower panel), for a configuration with
$q=3$, $\chi_1=0.5$, $\chi_2=0.25$. For this configuration, we show the quasi-circular dynamical quantities (red solid line), the eccentric quantities with initial eccentricity $e_0=0.3$
(solid green line) and the orbit-average curves (blue dashed line). 
The black dots, on the left of the orange vertical line, represent the  $\{\overline{t}_i, \overline{X}_i\}$ points used to construct the orbit-average curve, the orange vertical line represents the time $t^{\text{average-end}}=-30M$ from which the instantaneous quantities are used to construct the orbit-average curve, and the gray vertical line corresponds to time at which the inspiral ends, that is $t^{\text{inspiral-end}}$. The green dots, on the right of the vertical orange line, corresponds to values of the instantaneous eccentric quantities, which are used together with the black dots to generate the interpolated function representing the orbit-average curve.
The insets in both panels zoom into the last $1000M$ of evolution to
better show the behavior of the orbit average curves at the end of the
inspiral.
}
\label{fig:orbitAv}
\end{center}
\end{figure}

The  orbit averaging procedure  also requires the
introduction of boundary conditions at the start and at the end of the 
inspiral~\footnote{Once an orbit-average quantity, $\{\langle t\rangle_i, \langle
X\rangle_i\}$,  has been computed, it is interpolated using \texttt{cubic
splines} GSL routine \cite{galassi2018scientific} so that it can be evaluated
onto the time grid with a sampling rate corresponding to the one specified by the
user.}. At the start of the inspiral, we simply use the
the time of the first maximum, $t^\text{first-max}$. The impact of this choice is negligible because at this point 
of the evolution the orbit-average NQC function is quite smooth. Meanwhile, at the end of the
inspiral, we need to reproduce the plunging behavior of the dynamical quantities 
$r, \omega, p_{r_*}$ to accurately compute the NQC function, as done in the quasi-circular 
case. To achieve this goal,  we impose that from a certain time, $t^{\text{average-end}}$, the orbit-average 
dynamical quantities follow the non-orbit--average dynamics. In Fig.
\ref{fig:orbitAv}, we show how the orbit-average  $\omega$ and $p_{r_*}$ are constructed
from the instantaneous eccentric dynamics for a particular eccentric
configuration, $q=3,  \quad \chi_1=0.5, \quad \chi_2=0.25, \quad e_0=0.3$.
Additionally, the  evolution of the quasi-circular quantities is included in Fig.
\ref{fig:orbitAv}, showing that the orbit-average curves for $\omega$ and $p_{r_*}$ agree remarkably
well with the ones corresponding to the quasi-circular evolution.

\begin{figure}[htbp!]
\begin{center}
\includegraphics[width=1.\columnwidth]{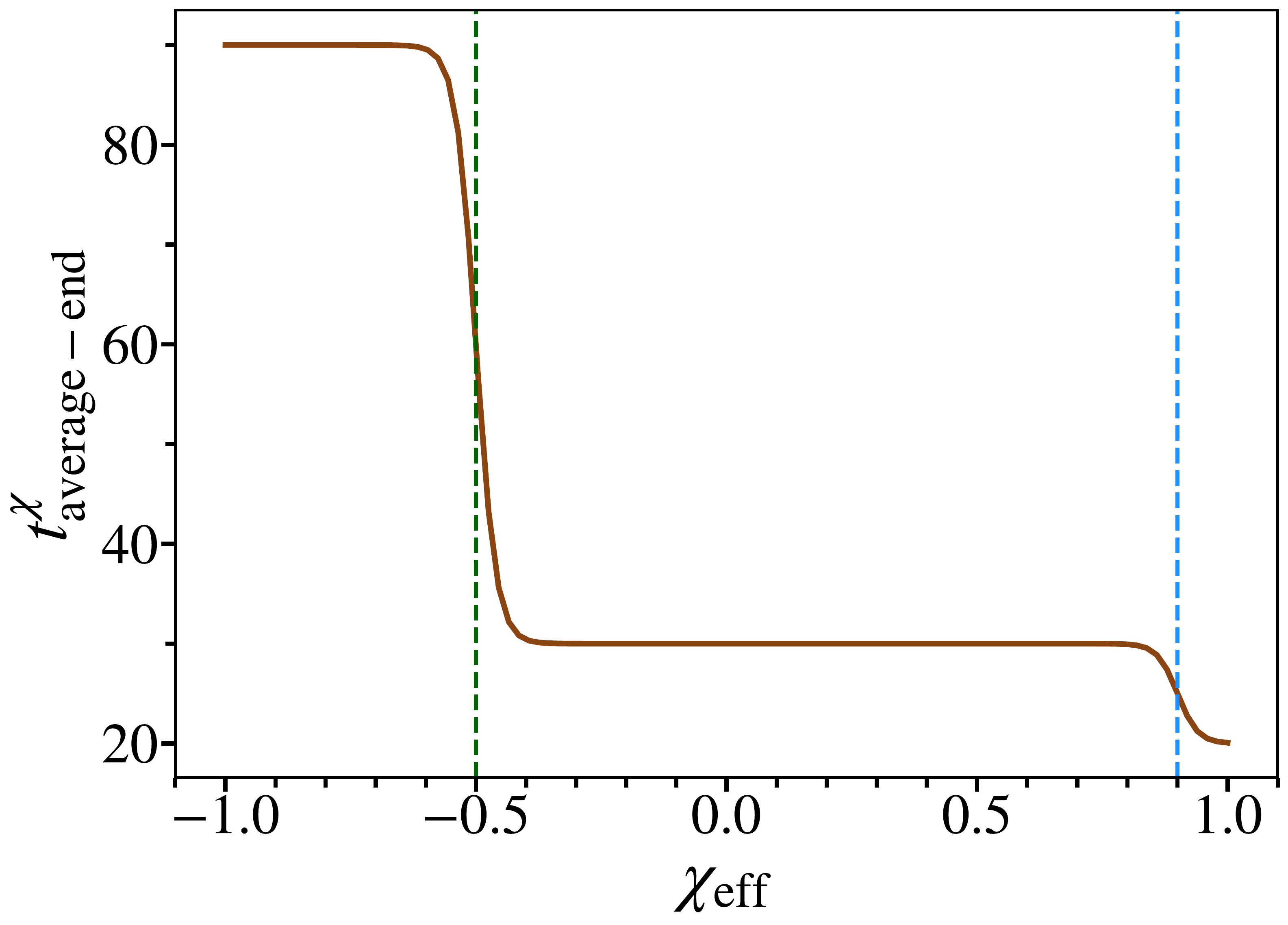}
\caption{Dependence of the $t_{\chi}^{\text{average-end}}$ parameter defined in Eq.
\eqref{eq:eq11.3} on the effective spin parameter. The green (blue) dashed
vertical line corresponds to $\chi_{\text{low}}=-0.5$
($\chi_{\text{high}}=0.95)$ values used to construct $t_{\chi}^{\text{average-end}}$.}
\label{fig:sigmoidFunction}
\end{center}
\end{figure}

In Fig. \ref{fig:orbitAv}, the orange vertical line corresponds to $t^{\text{average-end}}$, from which the values of the instantaneous, eccentric dynamical variables (green dots on the right of the orange vertical line) are attached, and together with the orbit-average values (black dots on the left of the orange vertical line) form the
orbit-average curve, which is then interpolated. Thus, between the time
of the last found maximum, $t^{\text{last-max}}$ (the closest black dot to the orange vertical line), and the value of $t^{\text{average-end}}$ there are no points to use. If the distance between $t^{\text{last-max}}$ 
and  $t^{\text{average-end}}$ is too large, one can get unphysical
oscillations coming from interpolation artefacts due to a too large interpolated
interval without points. By contrast if $t^{\text{average-end}}$ is very close to $t^{\text{last-max}}$
one could introduce residual oscillations due to eccentricity, or spurious oscillations due to the artefacts of the interpolation method, as at plunge the change of behavior of the dynamical quantities, especially $p_{r^*}$, challenges the interpolation procedure. Furthermore, we note that the position of $t^{\text{last-max}}$ depends substantially on the spins of the binary. For instance, for high-negative spins  $t^{\text{last-max}}$ typically occurs far from the merger ($t^{\omega_\text{peak}}- t^{\text{last-max}} \gtrsim 100M$), while for high-positive spins $t^{\text{last-min}}$ can be very close to merger  $t^{\omega_\text{peak}}- t^{\text{last-max}} \lesssim 50M$.
We use the following phenomenological prescription for the dependence
of  $t^{\text{average-end}}$ with spins,
\begin{equation}
\begin{split}
& t_{\chi}^{\text{average-end}}(t_j^{\text{average-end}},\chi_{\text{low}},\chi_\text{high}, \beta_\chi; \chi_{\text{eff}}) = \\ 
& \quad \qquad t^{\text{average-end}}_0 \times \left[ 1- w(\beta_{\chi},\chi_{\text{low}};\chi_{\text{eff}}) \right]\\
& \qquad +  t^{\text{average-end}}_1 \times \left[ 1- w(\beta_\chi,\chi_\text{high};\chi_{\text{eff}})\right]\\
& \qquad +  t_2^{\text{average-end}} \times w(\beta_\chi,\chi_{\text{high}};\chi_{\text{eff}}),
\end{split}
\label{eq:eq11.2}
\end{equation}
where  the function $w$ is the sigmoid defined in Eq. \eqref{eq:eq7}, and, the effective spin parameter is defined as,
\begin{equation}
\chi_{\text{eff}} = \frac{m_1 \chi_1 + m_2 \chi_2}{M}.
\label{eq:eq11.3}
\end{equation}
In Eq. \eqref{eq:eq11.2} the indices takes the values $j=0,1,2$, while $\beta_{\chi}=50$,
$\{\chi_{\text{low}},\chi_{\text{high}}\}=\{-0.5,0.95\}$ and
$(t^{\text{average-end}}_0,t^{\text{average-end}}_1,t^{\text{average-end}}_2 )= (60M, 30M,20M)$. In
Fig. \ref{fig:sigmoidFunction} the dependence of $t^{\chi}_{\text{average-end}}$  on the
effective spin parameter is illustrated. The values of the parameters
$\{t^{\text{average-end}}_0,t^{\text{average-end}}_1,t^{\text{average-end}}_2 ,
\chi_{\text{low}},\chi_{\text{high}}\}$ are chosen after evaluating the model in
a grid of points in  parameter space $q=[1-20]$,
$\chi_{\text{eff}}=[-0.99,0.99]$, $e=[0,0.3]$ with $t_{\text{append}}=[10,100]$,
and imposing that the frequency of the (2,2)-mode does not have oscillations above
$20\%$ with respect to the (2,2)-mode frequency of the quasi-circular
\texttt{SEOBNRv4} model in the last 100M prior to the peak of the (2,2)-mode
amplitude. We note that the current prescription for the $t^{\text{average-end}}$ parameter is independent of the mass 
ratio, because we find that the spin effects are dominant for the mass ratios we 
consider here.

We note that the  orbit averaging procedure  fails when eccentricity
  goes to zero due to the absence of maxima in the orbital quantities, 
  as they become non-oscillatory functions. Hence, in order to have a
  smooth transition to the quasi-circular limit, we need a new metric to measure the distance between the time of the last found maxima, $t^{\text{last-max}}$, and the time corresponding to  $t^{\text{average-end}}_\chi$. This is due to the fact that the smaller the eccentricity, the earlier $t^{\text{last-max}}$ occurs in the inspiral, and thus, the larger the region without points over which the orbit-average quantities would be interpolated. In practice, we find
  that one can use the time at which the inspiral ends\footnote{We refer to the last point of the evolution of the equations of motion given by Eqs. \eqref{eq:eq1}.}, $t^{\text{inspiral-end}}$ (dashed gray vertical line in
  Fig. \ref{fig:orbitAv}), 
 instead
  of $t^{\text{average-end}}_\chi$, because for low eccentricities a
  difference of $\sim 30-100M$ has negligible impact in the  orbit averaging procedure. Hence, we impose that $t^{\text{average-end}}$ depends on
\begin{equation}
\begin{split}
\Delta t = t^{\text{inspiral-end}} - t^{\text{last-max}},
\end{split}
\label{eq:eq11.4}
\end{equation}
such that the final expression for the $t^{\text{average-end}}$ reads as follows,
\begin{equation}
\begin{split}
t^{\text{average-end}}(\Delta t,\chi_{\text{eff}}) & =  t_{\chi}^{\text{average-end}}(\chi_{\text{eff}}) \times [1-w(\beta_t, \Delta t_0;\Delta t)] \\ 
& + \alpha_t \times w(\beta_t,  \Delta t_0; \Delta t) \times  \Delta t,
\end{split}
\label{eq:eq11.5}
\end{equation}
where $\beta_t =0.1$, $\Delta t_0=350M$ and  $\alpha_t=0.75$. This choice of
parameters ensures that $t^{\text{average-end}}$ increases as the time at which the last maximum is found occurs at earlier times in the inspiral for low-initial eccentricities, thus smoothly increasing the region in which the instantaneous variables are used to construct the orbit-average quantities. These values were set after testing and evaluating more than $10^4$ waveforms for initial
eccentricities $e<0.1$ in the parameter space  $q=[1,20]$,
$\chi_{\text{eff}}=[-0.9,0.9]$. Finally, we only consider the  orbit averaging procedure  when at least three maxima are found during the inspiral, otherwise we use the instantaneous dynamics to construct the NQC function in Eq. \eqref{eq:eq6}.


\section{PN expressions for dynamical quantities in the Keplerian parametrization} 
\label{app:eccICs}

In this appendix, we provide the expressions of the dynamical quantities $p_r,\dot{p}_r,$ and $\dot{r}$ in the Keplerian parametrization. They are needed for calculating the initial conditions for eccentric orbits, as discussed in Sec.~\ref{sec:InitialConditions}.

In the Keplerian parametrization we have:
\begin{equation}
\label{rKep}
r = \frac{1}{u_p (1 + e \cos \zeta)}\,,
\end{equation} 
where $u_p$ is the inverse semilatus rectum and $\zeta$ is the relativistic anomaly.
Inverting the Hamiltonian at periastron and apastron, $r_\pm = [u_p(1 \pm e)]^{-1}$, and solving for the energy $E$ and $u_p$ through to 2PN order, we obtain
\begin{widetext}
\begin{align}
\label{ELeup}
E &= \frac{e^2-1}{2 p_{\phi }^2}
+ \frac{1-e^2}{8c^2 p_{\phi }^4}\left[e^2 (\nu -7)-\nu -9\right]
+\frac{e^2-1}{16c^4 p_{\phi }^6} \left[e^4 (\nu^2 -7\nu  +33)-2 e^2 (\nu^2 +9\nu-71)+(\nu -7) \nu +81\right] \nonumber\\
&\quad
+\frac{1-e^4}{c^3 p_{\phi }^5} \left[2 \delta  \chi _A-(\nu -2) \chi _S\right]
+\frac{1-e^2}{2c^4L^6} \left\lbrace
\chi _S^2\left[e^2 \left(-8 \nu ^2+8 \nu -3\right)-1\right] 
+\chi _A^2\left(3 e^2+1\right) (4 \nu -1) 
+2 \delta  \chi _A \chi _S \left[e^2 (4 \nu -3)-1\right] \right\rbrace, \nonumber\\
u_p &= \frac{1}{p_{\phi }^2} 
+ \frac{e^2+3}{c^2p_{\phi }^4}
+\frac{e^2+3}{c^4 p_{\phi }^6} \left(2 e^2-\nu +6\right)
+\frac{e^2+3}{c^3 p_{\phi }^5} \left[-2 \delta  \chi _A+(\nu -2) \chi _S\right] \nonumber\\
&\quad
+\frac{2}{c^4 p_\phi^6} \left\lbrace
\chi _S^2\left[e^2 \left(3 \nu ^2-3 \nu +1\right)+\nu ^2-\nu +1\right] 
-\chi _A^2\left(e^2+1\right) (4 \nu -1) 
-\delta  \chi _A \chi _S \left[e^2 (3 \nu -2)+\nu -2\right] \right\rbrace.
\end{align}

Inverting the Hamiltonian to obtain $p_r(E,p_\phi,r)$, then substituting Eqs.~\eqref{rKep} and~\eqref{ELeup}, yields $p_r(p_\phi,e,\zeta)$, which is given by the 2PN expansion
\begin{align}
\label{pr2PN}
p_r &=\frac{e \sin \zeta }{p_{\phi }} + \frac{e \sin \zeta }{c^2 p_{\phi }^3} \left(e^2+e \cos \zeta +2\right) + \frac{e^2 \sin \zeta  (e-\cos \zeta ) \left[-2 \delta  \chi _A+(\nu -2) \chi _S\right]}{c^3 p_{\phi }^4}
+ \frac{e\sin\zeta}{4 c^4 p_\phi^5} \bigg\lbrace 
8 e^4+e^2 (43-10 \nu )-12 \nu +22 \nonumber\\
&\quad
+\left[8 e^3+e (12-20 \nu )\right] \cos\zeta +e^2 (3-6 \nu ) \cos (2 \zeta )
+\chi _A^2 \left[e^2 (8 \nu -2) \cos (2 \zeta )+e^2 (4-16 \nu )+e (48 \nu -12) \cos\zeta+24 \nu -6\right] \nonumber\\
&\quad
+\chi _S^2 \left[-2 e^2 (1-2 \nu )^2 \cos (2 \zeta )+e^2 (8 (\nu -1) \nu +4)+e (-40 (\nu -1) \nu -12) \cos\zeta -6 (1-2 \nu )^2\right] \nonumber\\
&\quad
-\delta  \chi _A \chi _S \left[e^2 (4-8 \nu ) \cos (2 \zeta )+8 e^2 (\nu -1)+e (24-40 \nu ) \cos\zeta -24 \nu +12\right]
\bigg\rbrace.
\end{align}
Substituting $r$ and $p_r$ from Eqs.~\eqref{rKep} and~\eqref{pr2PN} into $\dot p_r = -\partial \hat{H}_\text{EOB}/\partial r$, and expanding yields
\begin{align}
\label{prdot2PN}
\dot p_r &= \frac{e \cos \zeta (e \cos \zeta+1)^2}{p_{\phi }^4} -\frac{e (e \cos \zeta +1)^2 \left[\left(e^2 (\nu -5)-\nu -7\right) \cos \zeta +e (\cos (2 \zeta )+3)\right]}{2 c^2 p_{\phi }^6} +\frac{e (e \cos\zeta+1)^2  \left[-2 \delta  \chi _A+(\nu -2) \chi _S\right]}{2c^3 p_{\phi }^7} \nonumber\\
&\qquad
\times \left[6 \left(e^2+1\right) \cos \zeta-e (3 \cos (2 \zeta )+1)\right] 
+\frac{e (e \cos \zeta+1)^2 }{8 c^4 p_{\phi }^8} \bigg\lbrace 2 e \left[\left(e^2 (\nu -7)-7 \nu -41\right) \cos (2 \zeta )+3 e^2 (\nu -7)-6 e \nu  \cos (3 \zeta )+11 \nu -59\right] \nonumber\\
&\qquad
+\left[e^4 (3 \nu^2 -17\nu +55)-6 e^2 (\nu^2 +3\nu-39)+3 \nu^2 -\nu+95\right] \cos \zeta \bigg\rbrace \nonumber\\
&\quad
+\frac{e (e \cos \zeta +1)^2}{4 c^4p_{\phi }^8} \bigg\lbrace
2 \delta  \chi _A \chi _S \left[\left(e^2 (19-26 \nu )+12 \nu \right) \cos (\zeta )+ 3 e^2 (2 \nu -1) \cos (3 \zeta )+2e (11 \nu -7) \cos (2 \zeta )+2e (\nu -1)\right] \nonumber\\
&\qquad
+\chi _S^2 \left[\left(e^2 \left(52 \nu ^2-52 \nu +19\right)-24 (\nu -1) \nu \right) \cos (\zeta )-e \left(3 e (1-2 \nu )^2 \cos (3 \zeta )+2 \left(22 \nu ^2-22 \nu +7\right) \cos (2 \zeta )+4 \nu ^2-4 \nu +2\right)\right] \nonumber\\
&\qquad
+\chi _A^2 e (4 \nu -1) [-19 e \cos \zeta+3 e \cos (3 \zeta )+14 \cos (2 \zeta )+2]
\bigg\rbrace.
\end{align}
Similarly, for $\dot r = \partial \hat{H}_\text{EOB}/\partial p_r$, we obtain
\begin{align}
\label{rdot2PN}
\dot{r} &= \frac{e \sin \zeta }{p_{\phi }} 
+\frac{e \sin \zeta }{2 c^2 p_{\phi }^3} \left[e^2 (-(\nu -1))-6 e \cos \zeta +\nu -3\right]
+ \frac{e^2 \sin \zeta  (\cos \zeta - e)}{c^3 p_{\phi }^4} \left[2 \delta  \chi _A+(2-\nu) \chi _S\right] \nonumber\\
&\quad
+ \frac{e}{8 c^4p_{\phi }^5} \left\lbrace 2 e \left(3 e^2 (\nu -3)+11 \nu -29\right) \sin (2 \zeta )+3 e^2 (2 \nu +1) \sin (3 \zeta )+\left[e^4 (3 (\nu -3) \nu +7)+e^2 \left(-6 \nu ^2+4 \nu +25\right)+3\nu^2 +27\nu-81\right] \sin \zeta \right\rbrace \nonumber\\
&\quad
+ \frac{e}{2c^4 p_\phi^5} \bigg\lbrace 
\chi _A^2 \sin \zeta (4 \nu -1) \left[-3 e^2+2 e \cos \zeta+1\right]
+\chi _S^2 \sin \zeta \left[e^2 \left(8 \nu ^2-8 \nu +3\right)-2 e \left(2 \nu ^2-2 \nu +1\right) \cos \zeta-(1-2 \nu )^2\right]
 \nonumber\\
&\qquad
+2 \delta  \chi _A \chi _S \sin \zeta \left[e^2 (3-4 \nu )+2 e (\nu -1) \cos \zeta+2 \nu -1\right]
\bigg\rbrace.
\end{align}
\end{widetext}

\bibliography{biblio}

\end{document}

%% file: Tables/ecc_NR_table.tex
\begin{tabular}{ c c c c c c c c | c c c | c c c }
 &   \multicolumn{7}{c|}{Numerical-relativity simulations} & \multicolumn{3}{c|}{\texttt{SEOBNRv4E}} & \multicolumn{3}{c}{\texttt{TEOBResumSE}} \\
\hline 
\hline 
$\text{ID}$ & q & $\chi_1$ & $\chi_2$ & $\omega_{\text{orb,p}}$ & $e_{\omega_{\text{orb,p}}}$ & $\omega_{\text{22,p}}$ & $e_{\omega_{\text{22,p}}}$ & $\omega_{p}$ & $e_{\omega_p}$ & $1-\bar{\mathcal{F}}_{\text{max}}[\%]$ & $\omega_{p}$ & $e_{\omega_p}$ & $1-\bar{\mathcal{F}}_{\text{max}} [\%]$ \\
\hline 
\hline 
SXS:BBH:0089 & 1 & -0.5 & 0.0 & 0.0128 & 0.06 & 0.025 & 0.048 & 0.0123 & 0.064 & 0.15 & 0.0111 & 0.064 & 0.64 \\
SXS:BBH:0321 & 1 & 0.33 & -0.44 & 0.0204 & 0.06 & 0.04 & 0.05 & 0.0196 & 0.07 & 0.22 & 0.0176 & 0.067 & 0.67 \\
SXS:BBH:0322 & 1 & 0.33 & -0.44 & 0.0223 & 0.075 & 0.0434 & 0.061 & 0.0224 & 0.086 & 0.35 & 0.0198 & 0.085 & 0.63 \\
SXS:BBH:0323 & 1 & 0.33 & -0.44 & 0.0235 & 0.126 & 0.045 & 0.102 & 0.0226 & 0.143 & 0.24 & 0.022 & 0.143 & 0.72 \\
SXS:BBH:0324 & 1 & 0.33 & -0.44 & 0.0303 & 0.246 & 0.0554 & 0.172 & 0.0299 & 0.297 & 0.3 & 0.0287 & 0.286 & 0.8 \\
SXS:BBH:1136 & 1 & -0.75 & -0.75 & 0.0244 & 0.09 & 0.0475 & 0.076 & 0.0231 & 0.113 & 0.31 & 0.0209 & 0.113 & 0.24 \\
SXS:BBH:1149 & 3 & 0.7 & 0.6 & 0.0197 & 0.048 & 0.0385 & 0.037 & 0.0189 & 0.045 & 0.3 & 0.0164 & 0.046 & 0.27 \\
SXS:BBH:1169 & 3 & -0.7 & -0.6 & 0.0156 & 0.045 & 0.0306 & 0.037 & 0.016 & 0.046 & 0.39 & 0.0115 & 0.0 & 0.79 \\
SXS:BBH:1355 & 1 & 0.0 & 0.0 & 0.0216 & 0.073 & 0.0421 & 0.059 & 0.0208 & 0.086 & 0.23 & 0.0189 & 0.07 & 0.79 \\
SXS:BBH:1356 & 1 & 0.0 & 0.0 & 0.0182 & 0.127 & 0.0347 & 0.1 & 0.0179 & 0.145 & 0.24 & 0.0172 & 0.145 & 0.66 \\
SXS:BBH:1357 & 1 & 0.0 & 0.0 & 0.0238 & 0.139 & 0.0453 & 0.112 & 0.0231 & 0.162 & 0.24 & 0.0224 & 0.159 & 0.74 \\
SXS:BBH:1358 & 1 & 0.0 & 0.0 & 0.0243 & 0.137 & 0.0464 & 0.111 & 0.0237 & 0.164 & 0.14 & 0.0226 & 0.148 & 0.74 \\
SXS:BBH:1359 & 1 & 0.0 & 0.0 & 0.0247 & 0.136 & 0.0472 & 0.111 & 0.0238 & 0.158 & 0.17 & 0.0234 & 0.162 & 0.69 \\
SXS:BBH:1360 & 1 & 0.0 & 0.0 & 0.0278 & 0.192 & 0.0522 & 0.156 & 0.0272 & 0.232 & 0.2 & 0.0259 & 0.218 & 0.78 \\
SXS:BBH:1361 & 1 & 0.0 & 0.0 & 0.028 & 0.194 & 0.0528 & 0.162 & 0.0277 & 0.239 & 0.24 & 0.0262 & 0.221 & 0.7 \\
SXS:BBH:1362 & 1 & 0.0 & 0.0 & 0.0319 & 0.255 & 0.0584 & 0.193 & 0.0313 & 0.308 & 0.4 & 0.0305 & 0.309 & 0.73 \\
SXS:BBH:1363 & 1 & 0.0 & 0.0 & 0.0321 & 0.257 & 0.0596 & 0.221 & 0.0313 & 0.304 & 0.48 & 0.0305 & 0.307 & 0.74 \\
SXS:BBH:1364 & 2 & 0.0 & 0.0 & 0.0215 & 0.059 & 0.0421 & 0.048 & 0.0203 & 0.059 & 0.49 & 0.0185 & 0.066 & 0.46 \\
SXS:BBH:1365 & 2 & 0.0 & 0.0 & 0.0215 & 0.083 & 0.0418 & 0.067 & 0.021 & 0.101 & 0.25 & 0.0191 & 0.101 & 0.29 \\
SXS:BBH:1366 & 2 & 0.0 & 0.0 & 0.0239 & 0.134 & 0.0456 & 0.111 & 0.0233 & 0.158 & 0.2 & 0.0226 & 0.152 & 0.29 \\
SXS:BBH:1367 & 2 & 0.0 & 0.0 & 0.0251 & 0.125 & 0.048 & 0.102 & 0.028 & 0.126 & 0.5 & 0.0264 & 0.113 & 0.92 \\
SXS:BBH:1368 & 2 & 0.0 & 0.0 & 0.0244 & 0.132 & 0.0466 & 0.107 & 0.0236 & 0.151 & 0.32 & 0.0233 & 0.157 & 0.33 \\
SXS:BBH:1369 & 2 & 0.0 & 0.0 & 0.0309 & 0.257 & 0.0573 & 0.191 & 0.0305 & 0.304 & 0.35 & 0.0299 & 0.308 & 0.39 \\
SXS:BBH:1370 & 2 & 0.0 & 0.0 & 0.0315 & 0.25 & 0.0585 & 0.144 & 0.0312 & 0.301 & 0.31 & 0.0301 & 0.296 & 0.45 \\
SXS:BBH:1371 & 3 & 0.0 & 0.0 & 0.0213 & 0.078 & 0.0414 & 0.063 & 0.0191 & 0.101 & 0.16 & 0.0191 & 0.093 & 0.15 \\
SXS:BBH:1372 & 3 & 0.0 & 0.0 & 0.0237 & 0.13 & 0.0454 & 0.107 & 0.0233 & 0.153 & 0.19 & 0.0228 & 0.153 & 0.26 \\
SXS:BBH:1373 & 3 & 0.0 & 0.0 & 0.0239 & 0.129 & 0.0458 & 0.104 & 0.0234 & 0.151 & 0.29 & 0.0229 & 0.149 & 0.21 \\
SXS:BBH:1374 & 3 & 0.0 & 0.0 & 0.0306 & 0.248 & 0.0562 & 0.206 & 0.0304 & 0.294 & 0.33 & 0.0296 & 0.293 & 0.31 \\
\hline 
\hline 
\end{tabular}